\documentclass[fleqn,usenatbib]{mnras}

\usepackage{newtxtext,newtxmath}
\usepackage[T1]{fontenc}
\usepackage{ae,aecompl}
\usepackage[version=3]{mhchem}

\usepackage{amssymb,amsfonts,amsmath,amstext,amsgen,amsopn,amsxtra,indentfirst,times}%

\usepackage{savesym}%
\usepackage{graphicx}%
\usepackage{hyperref}%
\usepackage{lineno}%
\usepackage{mathtools}%
\usepackage{multirow}
\usepackage{amsmath}%
\usepackage{xspace}%
\usepackage{units}%
\usepackage{tikz}%
\usetikzlibrary{positioning}%
\usetikzlibrary{shapes,arrows}%
\usepackage{nicefrac}%
\usepackage{array}

\newcommand{\RE}{R$_{\rm \earth}$\xspace}
\newcommand{\ME}{M$_{\rm \earth}$\xspace}
\newcommand{\DE}{$\rho_{\rm \earth}$\xspace}

\newcommand\rev[1]{\textcolor{black}{#1}} % FOR CARO NOTES

\title[Ca-Al-rich Super-Earths]{A new class of Super-Earths formed from high-temperature condensates: HD219134~b, 55~Cnc~e, WASP-47 e}

\author[Dorn, Harrison, Bonsor, Hands]{
C. Dorn$^{1}$\thanks{E-mail: cdorn@physik.uzh.ch},
J. H. D. Harrison$^{2}$,
A. Bonsor$^{2}$,
T. O. Hands$^{1}$
\\
$^{1}$University of Zurich, Institut of Computational Sciences, Winterthurerstrasse 190, CH-8057, Zurich, Switzerland.  Emails: cdorn@physik.uzh.ch \\
$^{2}$University of Cambridge, Institute of Astronomy, Madingley Road, CB3 0HA Cambridge, United Kingdom
}

\date{Accepted XXX. Received YYY; in original form ZZZ}

\pubyear{2018}

\begin{document}
\sloppy
\label{firstpage}
\pagerange{\pageref{firstpage}--\pageref{lastpage}}
\maketitle
% * <dorn.caro@gmail.com> 2018-09-26T09:52:35.680Z:
\begin{abstract}
We hypothesise that differences in the temperatures at which the rocky material condensed out of the nebula gas can lead to differences in the composition of key rocky species (e.g., Fe, Mg, Si, Ca, Al, Na) and thus planet bulk density. Such differences in the observed bulk density of planets may occur as a function of radial location and time of planet formation. In this work we show that the predicted differences are on the cusp of being detectable with current instrumentation. In fact, for HD 219134, the 10 \% lower bulk density of planet b compared to planet c could be explained by enhancements in Ca, Al rich minerals. However, we also show that the 11 \% uncertainties on the individual bulk densities are not sufficiently accurate to exclude the absence of a density difference as well as differences in volatile layers. Besides HD 219134 b, we demonstrate that 55 Cnc e and WASP-47 e are similar candidates of a new Super-Earth class that have no core and are rich in Ca and Al minerals which are among the first solids that condense from a cooling proto-planetary disc. Planets of this class have densities 10-20\% lower than Earth-like compositions and may have very different interior dynamics, outgassing histories and magnetic fields compared to the majority of Super-Earths.

\end{abstract}


%%%%%%%%%%%%%%%%%%%%%%%%%%%%%%%%%%%%%%%%%%%%%%%%%%

%%%%%%%%%%%%%%%%% BODY OF PAPER %%%%%%%%%%%%%%%%%%

\section{Introduction}
%intro from amy
Rocky planets form out of the solid bodies leftover when the proto-planetary gas disc (PPD) disperses. In the inner regions of PPDs these solids condense out of the nebula gas as the disc cools.  At temperatures higher than 1200 K, if the condensates form in chemical equilibrium, large compositional differences in terms of key refractory elements such as Fe, Mg, Si, Ca, Al, Ti, and Na can occur \citep[e.g.,][]{Lodders2003}. Compositional differences can lead to differences in the bulk density of rocky planets. Here, we investigate the possible variability in planet bulk density that is due to the chemical variability as inherited from planetesimals formed at different temperatures.

%solar system from amy
In our Solar System, trends in the depletion of chondritic meteorites in moderately volatile species as a function of their estimated radial origin and compared to bulk Earth, highlight the importance of differences in condensation temperature of planet building compounds. The importance of radial migration or mixing in the PPD, however and how these \rev{influence compositional variability of planet building blocks} is not fully understood \citep{gail2004radial}. 

Chemical and dynamical disc processes influence the distribution of observed exoplanets. For terrestrial planets, relative abundances of major rock-forming elements control their bulk composition. For example, Mg/Si govern the distribution of different silicates, while C/O control the amount of carbides versus silicates. How elemental ratios within the PPD vary as a function of time and radial distance and how that affects structures and compositions of formed planets is a subject of ongoing research. 

In general, planets that form within the same PPD can have very different volatile element budgets \citep[e.g.,][]{oberg2016excess} but have generally similar budgets in relative refractory elements \citep[e.g.,][]{elser2012origin,bond2010making,thiabaud2014stellar,sotin2007mass}. 
The reason is that the disc region where condensation temperatures of different volatile compounds ($<200$K) are reached is very extended (semi-major distances $a>1$AU). For refractory compounds, this region is limited. Here, refractory elements are Al, Ca, Mg, Si, Fe, Na that are rock-forming compounds, while volatile elements include S, C, O, N, He, H. Only very close to the stars, the gas cools slowly such that not all refractory compounds can condense out before the gas disc disperses. Thus, except for this innermost region, the refractory element ratios of formed planetesimals are directly correlated to the PPD bulk composition, which in turn is commonly assumed to be represented by the host star composition.     % ($\Delta a = 0.2-1.7$AU, depending on the age of the disc). 
How refractory elemental ratios of a formed planet can deviate from its host star chemistry has been investigated for Earth given the volatility trends of various elements in the solar nebular by \citet{wang2018elemental} with applications to exoplanets \citep{wang2018enhanced}.
The majority of rocky planets from the same system follow the same mass-radius trend. For the Solar System, this is in fact the case for Mars, Venus, and Earth. Although their atmospheres are very different in mass and chemistry, their atmospheres contribute little to planet radii. 
%Although following the same mass-radius trend, there are differences measured between Earth rocks, Martian and lunar meteorites. The differences in isotope chemistry can inform the time of terrestrial planet formation \citep{dauphas2011hf,morbidelli2012building}. 

% why can there be deviations from m-r-relationship
Observed Super-Earths indicate an inherent scatter in bulk densities indicative of variable interior compositions and structures. Even if a Super-Earth would directly inherit the relative abundances of refractory elements from its star, deviations in bulk density are  generally possible by various mechanisms.
Deviations towards higher densities can be caused by giant impacts \citep{benz1988} or tidal disruption \citep{rappaport2013}. Deviations towards lower densities are usually due to different budgets in volatile-rich layers (gas or ice). Here, we also discuss the possibility of different rock composition as inherited from planetesimals formed at different temperatures.  Likelihood and magnitude of the deviations are individual to each scenario. Here, we attempt to qualitatively and quantitatively discuss the probability of each scenario for the characterization of the interiors of HD219134 b and c, as well as 55 Cnc e and WASP-47 e. 

%HD 219134
The two rocky planets in the K-dwarf system HD219134 \citep{gillon2017two} are curious in that they do not follow the same mass-radius trend, but show a 10\% density difference.
%alternative scenarios
\citet{dorn2018secondary} have shown that their interiors can be explained by using stellar abundance constraints on refractory elements. The lower bulk density of planet b was suggested to be due to a secondary atmosphere. Besides a possible difference in atmospheric thicknesses, we will discuss here the above-mentioned difference in rock composition. Alternatively, the observed difference in density may not be real, given uncertainties in radius and mass determinations. 

%55Cnc-e
For the highly irradiated Super-Earth 55 Cnc e, numerous interior characterization studies aim to explain its relatively low density 
%$\rho_{\rm 55Cnce} = 1.3 \pm 0.08 $\DE 
and its variable nature \citep[e.g.,][]{demory2015variability,demory2016heatmap,demory2016a,angelo2017case,crida2018mass,bourrier201855} however the nature of this planet remains inconclusive. 
WASP-47 e is similar to 55 Cnc e in that it is on an ultra-short orbit and has a density that is too low for a rocky Earth-like composition \citep{vanderburg2017precise}. One possible explanation for the low densities is that these planets are remnant cores of hot Jupiters in the state of gas loss \citep{valsecchi2014hot}. However, this mechanism  cannot explain the general population of planets on ultra-short orbits (USPs), plus an escape of hydrogen from 55 Cnc e was not detected \citep{ehrenreich2012hint}.
\citet{vanderburg2017precise} highlight that both well-characterized planets WASP-47 e and 55 Cnc e are similar and not typical for USPs and may require a more exotic origin compared to other rocky USPs.

The paper is structured as follows. First, we provide a detailed analysis of the interiors of HD219134 b and c. We discuss the probability of the bulk densities being caused by a difference in bulk rock composition (Section \ref{Sec:refr}),  a difference in volatile layers (Section \ref{Sec:vol}) or due to observational biases (Section \ref{Sec:bias}). We then discuss the interiors of 55 Cnc e (Section \ref{Sec:55cnc}) and WASP-47 e (Section \ref{Sec:wasp47}) and propose how to find further candidates (Section \ref{Sec:further}). We finish with conclusions in Section \ref{conclusions}.

\section{HD219134 b and c}
\subsection{Previous studies on HD219134 b and c}
%what has been previously discussed about their possible interiors?
Given the bulk densities of HD219134 b and c (Table \ref{tab:planetdata}), \citet{gillon2017two} suggested purely rocky interiors and relate the density difference to different core mass fractions (planet b: $0.09^{+0.16}_{-0.09}$, planet c: and $0.26\pm 0.17$). Their different core mass fractions imply different bulk rock compositions, which is difficult to explain other than by compositional variability of building blocks from the disc, which we provide here. 

\citep{gillon2017two} also consider the possibility of a thick H-dominated atmosphere and/or water layers to explain planet b's lower density. Considering evaporative loss, \citet{dorn2018secondary} conclude that the possible atmospheres are unlikely to be dominated by H but gas of heavier mean molecular weight, i.e., outgassed from the interior.

\begin{figure*}
	\includegraphics[width=1.6\columnwidth]{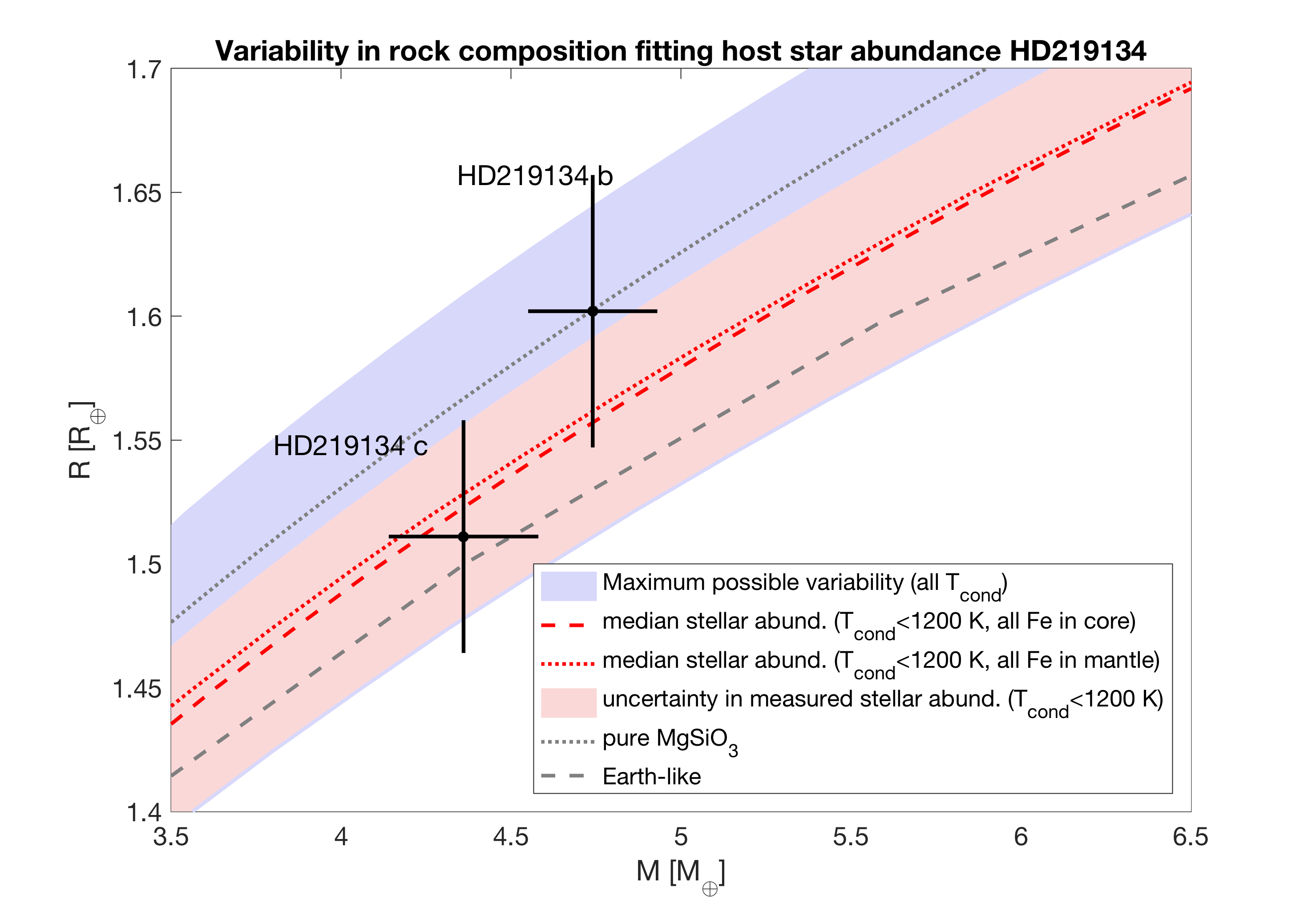}
    \caption{Mass-radius plot showing HD219134 b and c compared to scaled idealized interiors. We assume a 1:1 compositional relationship (i.e., ratios of refractory elements) between the host star HD219134 and the PPD. The variability of purely rocky planets forming at different times and locations within the disc is highlighted by the blue area.
    Purely rocky interiors that are built from temperate condensates ($T_{\rm cond}<1200$K) lie within the red area, which respects the uncertainty in measured stellar abundances.
    Interiors that fit the median stellar abundance follow the red curves. The difference between the dashed and dotted red curves is due to the degeneracy between iron-free mantles with pure iron cores and core-free interiors with all iron in the mantle, respectively.
    The light red area accounts for uncertainty in abundance estimates.}
    \label{fig:MR_curve}
\end{figure*}

The interior characterization by \citet{dorn2018secondary} used constraints on the relative refractory element ratios of the bulk planet as measured in the stellar photosphere \citet{dorn2015can}, assuming a direct correlation in-between. In fact, mass and radius of planet c can be well explained by a rocky interior that fit the stellar abundance constraint as illustrated in Figure \ref{fig:MR_curve}. Purely rocky interiors that fit the median stellar abundances follow the red curves, while the red area illustrate the associated uncertainty.

\begin{table*}
	\centering
	\caption{\rev{Data of the inner four planets of HD219134, for which the planets b, c, f, d, g, and h are known.}}
	\label{tab:planetdata}
\begin{tabular}{c c c c c}
\hline
parameter & planet b & planet c & planet f & planet d\\
% planet &  M [\ME] &R [\RE]  & $\rho$ [\DE] & $D$ [AU]\\
\hline
% b & 4.74 $\pm$ 0.19 & 1.602 $\pm$ 0.055 &1.15 $\pm$ 0.13 & 0.03876 $\pm$0.00047\\
% c & 4.36 $\pm$ 0.22 & 1.511 $\pm$ 0.047 &1.26 $\pm$ 0.14 & 0.06530 $\pm$0.00080\\
M [\ME] &4.74 $\pm$ 0.19  & 4.36 $\pm$ 0.22 & $> 7.30 \pm 0.40$ &$> 16.17 \pm 0.64$\\
R [\RE] &1.602 $\pm$ 0.055&1.511 $\pm$ 0.047 & $> 1.31 \pm 0.02$ &$> 1.61 \pm 0.02$\\
$\rho$ [\DE] &1.15 $\pm$ 0.13 &1.26 $\pm$ 0.14 &-- & --\\
$a$ [AU] & 0.03876 $\pm$0.00047& 0.06530 $\pm$0.00080 & 0.1463 $\pm$ 0.0018 & 0.237 $\pm$ 0.003\\
\hline
\end{tabular}
\end{table*}

The abundances for the nearby star HD~219134 were measured by a total of 9 groups within the literature \citep[e.g.,][]{thevenin1998chemical,prieto2004s,luck2005stars,valenti2005spectroscopic,mishenina2013abundances,ramirez2013oxygen,maldonado2015searching,da2015homogeneous}. Table \ref{tab:stardata} lists the median stellar refractory abundances from the \emph{Hypatia catalog} \citep{hinkel2014stellar} after outliers were removed. Outliers are those that lie beyond the range of possible abundances in stars with metallicities similar to HD219134 based on \citet{brewer2016c}.
The C/O ratio of HD~219134 is assumed to be 0.62, as this is the value found when the outliers are removed the \emph{Hypatia catalog} \citep{hinkel2014stellar}.  If the actual C/O ratio of  HD~219134 was outside the range of 0.25-0.75, our calculated disc chemistry could significantly differ. However, a recent statistical analysis from \citet{brewer2016c} showed that FGK stars cluster around slightly sub-solar C/O ratios of 0.47 and no super-solar C/O ratios of 0.7 were detected among the 849 sample stars.

\setlength{\tabcolsep}{6pt}
\begin{table}
\caption{Median stellar abundances of HD~219134 from Hypatia Catalog after the outliers and duplicate studies were removed. The unit is dex. The range of estimates from all the different studies is stated in brackets. \label{tab:stardata}}
\begin{center}
\begin{tabular}{ll}
\hline\noalign{\smallskip}
parameter & HD~219134 \\
\hline\noalign{\smallskip}
$[\rm Fe/H]$ & 0.09 (0.04 - 0.16)\\
$[\rm Mg/H]$ & 0.105 (0.09 - 0.16)\\
$[\rm Si/H]$& 0.055 (-0.03 - 0.12)\\
$[\rm Na/H]$& 0.19 (0.17 - 0.22)\\
$[\rm Al/H]$& 0.28 (0.24 - 0.29)\\
$[\rm Ca/H]$& 0.13 (0.09 - 0.18)\\
\hline
\end{tabular}
\end{center}
\end{table}

%%%%%%%%%%%%%%%%%%%%%%%%%%%%%%%%%%%%%%%%%%%%%%%%%%%%%%%%%%%%
%%%%%%%%%%%%%%%%%%%%%%%%%%%%%%%%%%%%%%%%%%%%%%%%%%%%%%%%%%%%
%%%%%%%%%%%%%%%%%%%%%%%%%%%%%%%%%%%%%%%%%%%%%%%%%%%%%%%%%%%%

\subsection{Different refractory element budgets as a cause for lower bulk density of planet b}
\label{Sec:refr}

In this Section, we investigate whether the density difference between planet HD~219134b and planet HD~219134c can be explained by different rock composition as inherited from the chemical heterogeneity of planetesimals formed at different temperatures.

\subsubsection{Planetesimal composition model}
\label{sec:HarrisonModel}
\begin{figure*}
	\includegraphics[width=.8\linewidth]{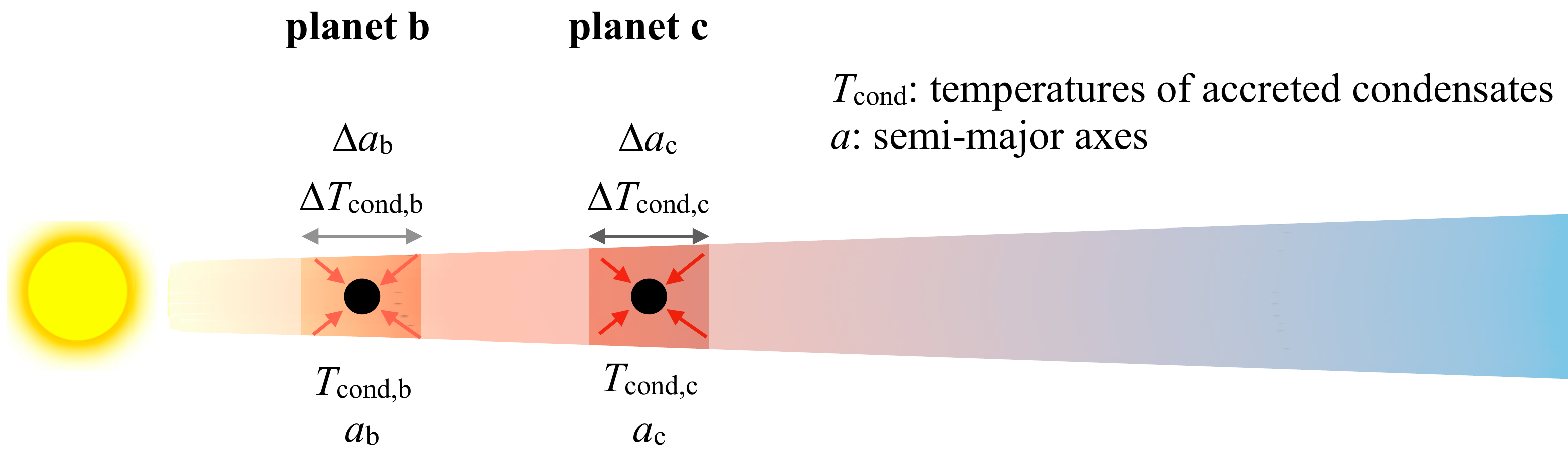}
    \caption{Schematic illustrating the growth of planets by accretion of condensates formed at different temperatures. The material that a planet at semi-major axis $a$ accretes from a feeding zone of width $\Delta a$ are condensates formed within the temperature range of $T_{\rm cond} \pm \nicefrac{1}{2} \Delta T_{\rm cond}$.   }
    \label{fig:sketch}
\end{figure*}

%introduction to the planetesimal compositional model
In order to model the bulk composition of the rocky planets HD~219134b and HD~219134c we employ a simple model which has been shown to recreate, to first order, the bulk composition of the rocky bodies in the Solar System \citep{Moriarty2014,harrison2018polluted}. The model assumes that rocky planets form via the aggregation of rocky planetesimals which have condensed out of a PPD in chemical equilibrium. The composition of the PPD is assumed to be identical to the stellar nebula, whilst the compositions of the planetesimals are determined by the compositions of the solid species found when minimising the Gibbs free energy at the pressures and temperatures present in the mid-plane of the PPD. In order to compare these compositions to that of the planets, we consider that the planets would form out of material that condensed out of the nebula within a small feeding zone around the planets orbital radii. Thus, the bulk compositions found are functions of the size of the feeding zone from which the planet accreted planetesimals ($\Delta a$), the time when the planetesimals condensed out of the disc ($t_{\rm disc}$), and the distance from the star at which the planet formed ($a$) (see Figure \ref{fig:sketch}). The compositions predicted by the model also depend on the mass and the composition of the host star ($M_*$ and $[X/H]_*$).

\begin{figure}
	\includegraphics[width=\columnwidth]{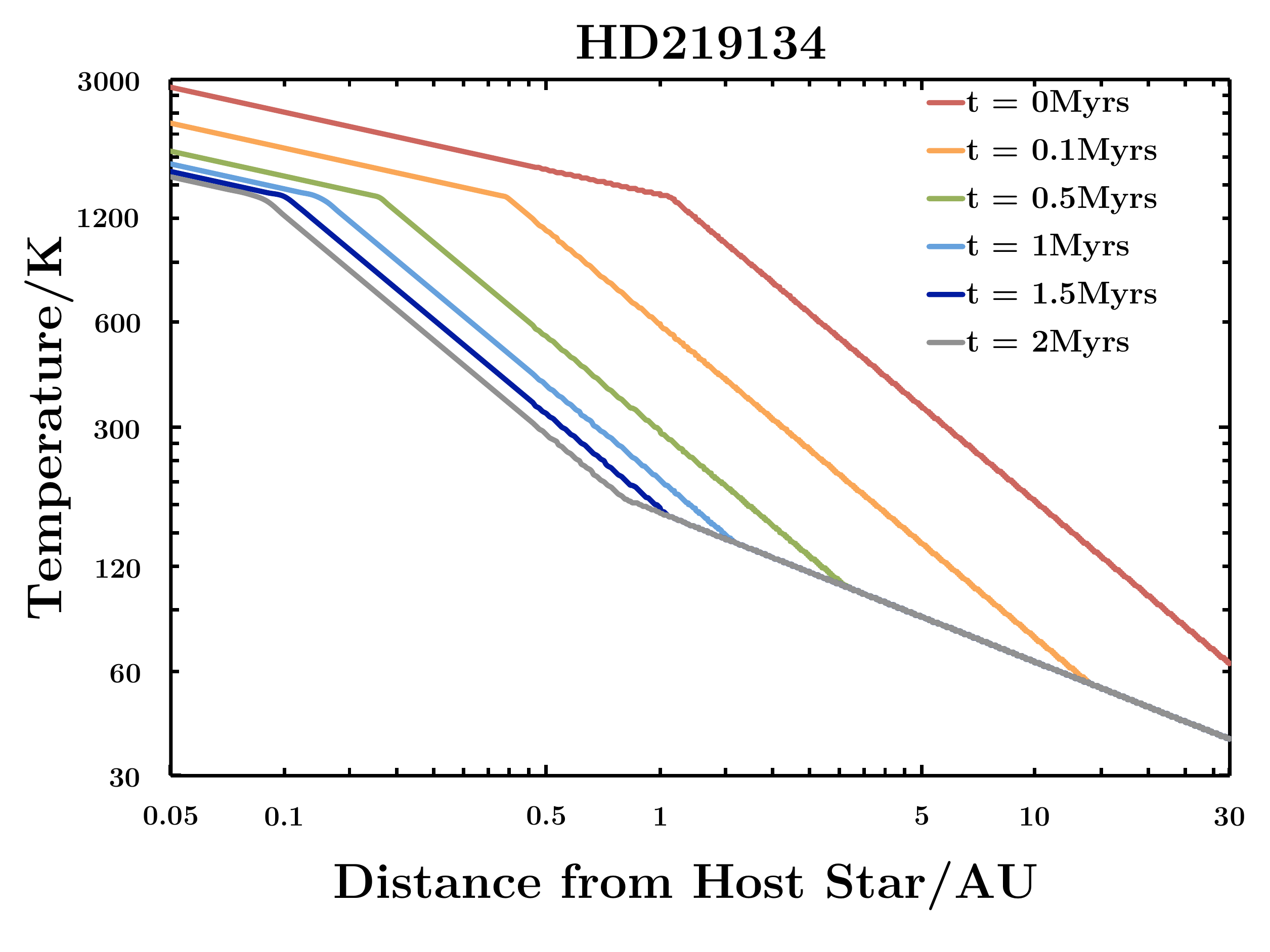}
    \caption{The modelled temperature in the mid-plane of the PPD of a HD219134-like star as a function of distance from the star ($a$) and time ($t_{\rm disc}$).}
    \label{fig:HDTR}
\end{figure}

\paragraph{Viscous irradiated PPD model}
\label{sec:disc}
%introduction and caveats of the disc model
The Gibbs free energy of the system, and thus the composition of the solids formed depends on the pressure and temperature at which condensation occurs. In order to consider reasonable pressures and temperatures for the inner regions of the PPD, and in order to convert these temperatures and pressures into radial locations within the disc and formation times for the solid condensates, we consider the simplest possible PPD model. We use the theoretical model derived in \cite{Chambers2009}, which models the viscous accretion of gas heated by the star. This model has been previously used for the modeling of planetesimal formation in PPDs \citep{Moriarty2014, harrison2018polluted} and super-Earths \citep{alessi2016formation}. This model ignores any vertical or radial mixing, and as will be discussed further later, any radial drift. All of these processes may be of critical importance in a  realistic PPD.
%outline of the values of the disc model used and the input constants required
The Chambers model is a disc model with an alpha parameterization which divides the disc into 3 sections; an inner viscous evaporating region, an intermediate viscous region, and an outer irradiated region. For the calculations in this work we have assumed disc parameters of $s_{0} = 33\,AU$, $\kappa_{0} = 0.3\,m^{2}kg^{-1}$ , $\alpha = 0.01$ , $\gamma = 1.7$, $\mu = 2.4$, and $M_{*}=0.78M_{\odot}$ following \citet{Chambers2009,motalebi2015}. We also assume that the mass of the PPD is directly proportional to the mass of the host star according to $M_{0}= 0.1M_{*}$ \citep{Chambers2009, Andrews2013}. The temperature and radius of the star in the PPD phase are assumed to be functions of the stellar mass in the form derived in \cite{siess2000}. The relations used in this work to calculate the PPD mass, the initial stellar radius, and the initial stellar temperature as a function of stellar mass are consistent with the values given in \cite{Stepinski1998} and \cite{Chambers2009} for a solar mass star.
% explanation and outline of the relevant equations in which the input constants are used
The analytical expressions for the pressure and temperature of the mid-plane of the PPD as a function of radial location ($a$) and time ($t_{\rm disc}$) are presented in Appendix \ref{discmodel}.
% Figures highlighting the PT space as a function of R and t for a Solar like and a HD219134 like disc
The temperature-radial location curves for the model disc around a star similar to HD~219134 are plotted as a function of time in Figure \ref{fig:HDTR}. The pressure-temperature space mapped out by the model disc for the case of a star similar to HD~219134 is displayed in the Appendix in Figure \ref{fig:HDPT}. The pressure-temperature space for the model disc of a solar mass star shows negligible differences compared to the HD~219134 case.

\paragraph{Equilibrium chemistry condensation model}
\label{sec:chem}
%introduction to the Chemistry model
We use the commercial Gibbs free energy minimisation package in HSC Chemistry version 8 to model the compositions of the solid species at the pressures and temperatures expected to be present in the PPD (\ref{sec:disc}). As the pressures and temperatures in the disc are a function of the formation time and the radial location ($t_{\rm disc}$, $a$), so are the planetesimal compositions.
% inputs and setup of the chemistry model
HSC chemistry version 8 was set up in the same way as \cite{bond2010making}, \cite{Moriarty2014}, and \cite{harrison2018polluted} which all used the software to model planetesimal compositions. The gaseous elements inputted, the list of gaseous species included in the model, the list of solid species included in the model, and the initial inputted gaseous abundances for the case of the HD~219134 system are displayed in Table \ref{tab:3}, Table \ref{tab:4}, Table \ref{tab:5}, and Table \ref{tab:6} respectively.
% tables outlining the setup of the HSC chemistry software  
\begin{table}
	\centering
	\caption{The gaseous elements which were included in the equilibrium chemistry code, HSC chemistry v. 8.}
	\label{tab:3}
\begin{tabular}{ c c c c c c c c}
\hline
\multicolumn{8}{c}{Gaseous Elements Included}\\
\hline
H & He & C & N & O & Na & Mg & Al \\
Si & P & S & Ca & Ti & Cr & Fe & Ni \\
\hline
\end{tabular}
\end{table}

\begin{table}
	\centering
	\caption{The list of possible gaseous species which could form in the equilibrium chemistry code, HSC chemistry v. 8.}
	\label{tab:4}
\begin{tabular}{ c c c c c c c c  }
\hline
\multicolumn{8}{c}{Gaseous Species Included}\\
\hline
Al     &   CrO &        MgOH &        PN  &       AlH  &      CrOH   &      MgS  &      PO                 \\
NS    &    SO       &    \ce{ CH4}      &    FeS     &     Na      &    SO2      &      CN    &    HC        \\             
Ca    &   HPO      &     NiH     &      SiP    &     CaH     &     HS     &        Cr   &     MgH       \\
 P     &   \ce{TiO2}    &       CrN    &     MgO    &    CaS    &      Mg    &         O  &        TiN          \\            
 CrS &       C  &      FeOH &     \ce{ H2O}    &   Ni     &    SiO &     TiO &     CrH    \\
\ce{N2}   &   \ce{Al2O}     &    AlOH    &     FeH    &   \ce{ NH3}      &   \ce{ S2}        &    \ce{ Na2}     &   Si           \\             
\ce{CO2} &  HCN   &        NaO     &     SiH   &      NiO     &    \ce{SiP2}      &    CaO      & \ce{H2S}         \\        
NiS     &  Ti        &       PH    &        TiS     &    AlS   &       FeO       &    NO    &    SN        \\
 PS   &      Fe    &       S &     \ce{H2}   &    NaH     &   SiC &  SiS  &    CaOH       \\
 HCO &    NaOH  &    SiN &     AlO &     S &   \ce{O2}   &     N &    MgN   \\
CO &   NiOH  &     CP &  He & & & & \\
\hline
\end{tabular}
\end{table}

\begin{table}
	\centering
	\caption{The list of possible solid species which could form in the equilibrium chemistry code, HSC chemistry v. 8.}
	\label{tab:5}
\begin{tabular}{ c c c c  }
\hline
\multicolumn{4}{c}{Solid Species Included}\\
\hline
\ce{Al2O3} &     \ce{ FeSiO3} &         \ce{CaAl2Si2O8} &           C         \\
SiC       &          \ce{Ti2O3}       & \ce{Fe3C}  &          \ce{Cr2FeO4}         \\
\ce{Ca3(PO4)2}  &     TiN          &      \ce{Ca2Al2SiO7}     &      Ni          \\
P       &               \ce{Fe3O4}          &         CaS         &            Si      \\
\ce{MgSiO3}      &       Cr         &              \ce{H2O}        &      \ce{CaMgSi2O6}     \\
  \ce{Fe3P}        &          \ce{CaTiO3}          &              Fe &                  AlN       \\
 \ce{MgAl2O4} &           \ce{Mg3Si2O5(OH)4}   &         MgS  &         \ce{CaAl12O19}  \\
        TiC      &               FeS   &      \ce{Mg2SiO4} &      \ce{Fe2SiO4} \\
    \ce{NaAlSi3O8} & \ce{NaAlO2} & \ce{Na2SiO3} & \\
\hline
\end{tabular}
\end{table}

\begin{table}
	\centering
	\caption{The inputted gaseous elemental abundances, the values are in kmol and are representative of the initial stellar nebula of HD219134 (Table \ref{tab:stardata}, \protect\citep{hinkel2014stellar}).}
	\label{tab:6}
\begin{tabular}{ c c   }
\hline
Element & Input \\
\hline
Al & $4.46 \times 10^{6}$ \\
C &  $3.71 \times 10^{8}$  \\
Ca &  $2.75 \times 10^{6}$ \\
Cr &  $6.17 \times 10^{5}$ \\
Fe &  $3.47 \times 10^{7}$ \\
H &  $1.00 \times 10^{12}$ \\
He &  $8.51 \times 10^{10}$ \\
Mg &  $4.32 \times 10^{7}$ \\
N &  $7.42 \times 10^{7}$ \\
Na &  $2.29 \times 10^{6}$ \\
Ni &  $2.19 \times 10^{6}$ \\
O &  $6.02 \times 10^{8}$ \\
P &  $2.82 \times 10^{5}$ \\
S &  $1.62 \times 10^{7}$ \\
Si &  $3.68 \times 10^{7}$ \\
Ti &  $1.41 \times 10^{5}$ \\
\hline
\end{tabular}
\end{table}
% outline of some of the justifications and caveats of the chemistry model
A caveat to the model is that the Gibbs free energy minimisation is only performed on the limited list of species outlined in Table \ref{tab:3}, Table \ref{tab:4}, and Table \ref{tab:5}, however, as these elements and species are the most abundant in the rocky debris in the solar system this is not thought to be a major limitation. The only major species missing from the list, that are expected to possibly alter the results, are the complex carbon macromolecules which are found in many asteroids and meteorites \citep{pizzarello2006} and whose formation mechanism is not yet understood. However, unless the carbon abundance in the disc is sufficiently super solar with respect to the overall metal abundances, these molecules will be trace species and therefore their contribution to the overall composition will be negligible.

\begin{figure}
	\includegraphics[width=\columnwidth]{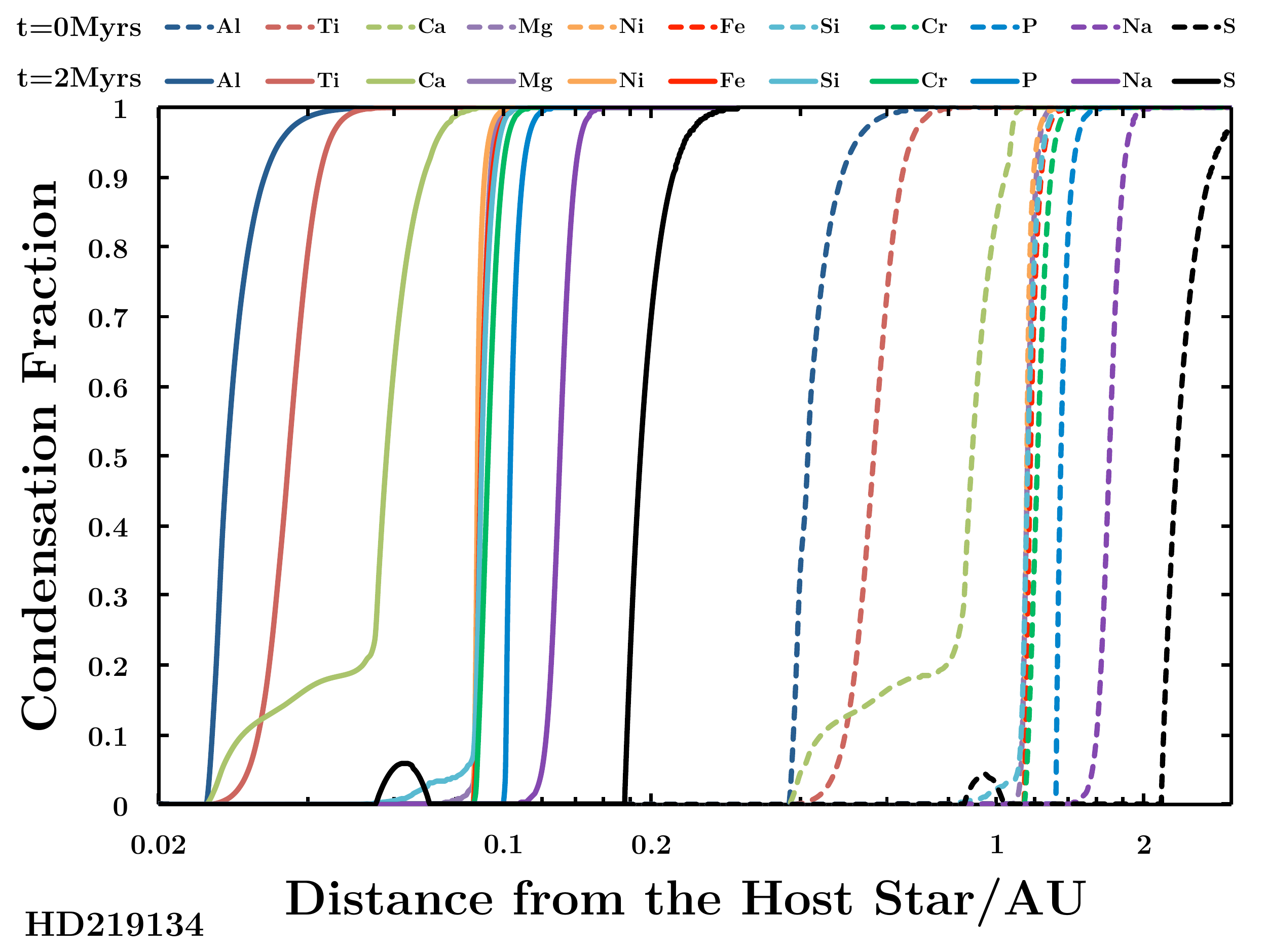}
    \caption{The ratio of solid species to gaseous species of each element as a function of radial location in the PPD at $t_{\rm disc}=0$\,Myrs and $t_{\rm disc}=2$\,Myrs for a HD~219134-like composition and disc model.}
    \label{fig:HDCD}
\end{figure}

\begin{figure}
	\includegraphics[width=\columnwidth]{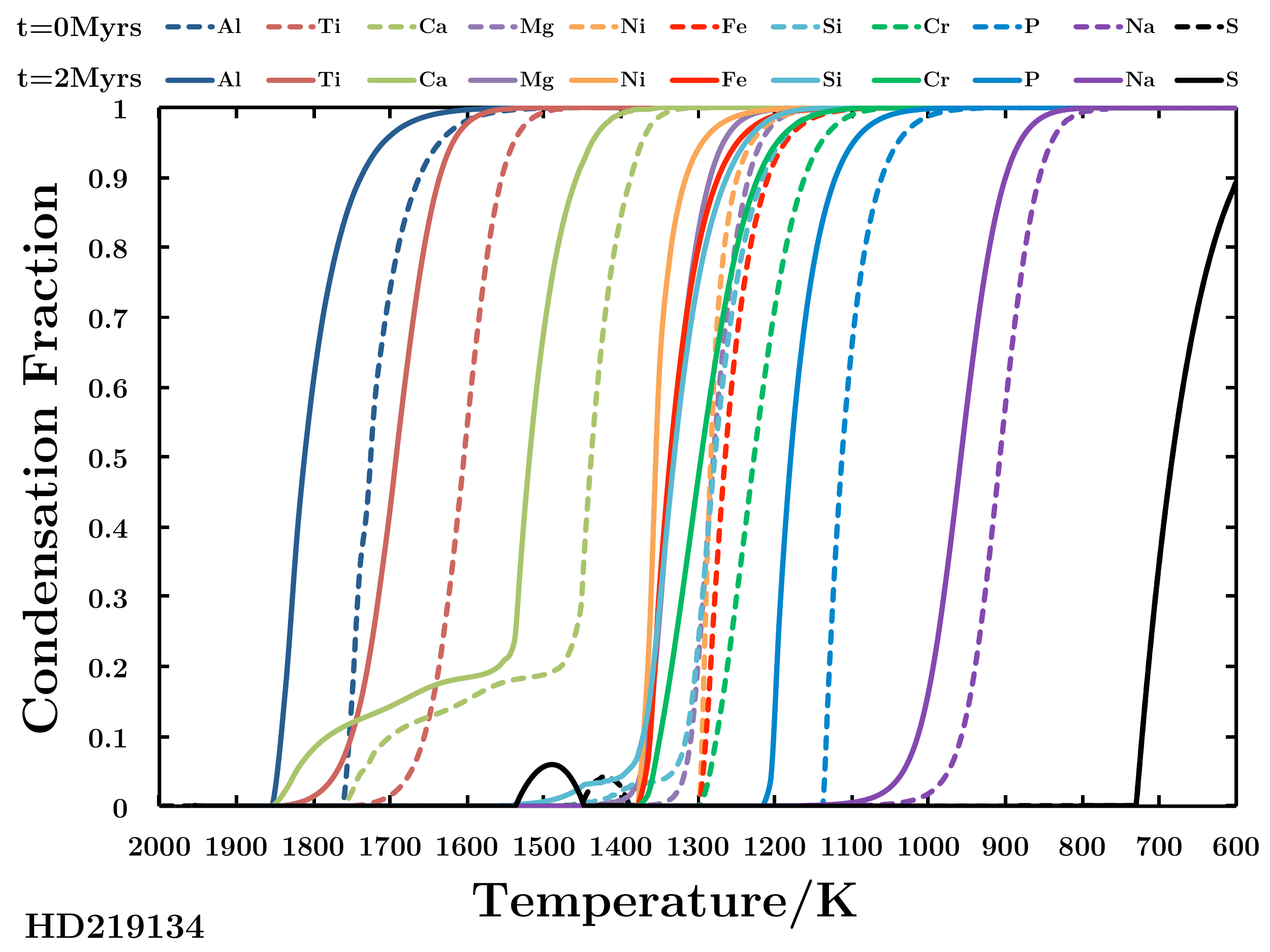}
    \caption{The ratio of solid species to gaseous species of each element as a function of temperature in the mid-plane of the PPD at $t_{\rm disc}=0$\,Myrs and $t_{\rm disc}=2$\,Myrs for a HD~219134-like composition and disc model.}
    \label{fig:HDCT}
\end{figure}

%Figures outlining the results of the chemistry model
Figure \ref{fig:HDCD} shows how the ratio of each element in solid state relative to gaseous state changes with increasing radial separation ($a$) from the host star at the two extremes of formation time ($t_{\rm disc}=0$\,Myrs and $t_{\rm disc}=2$\,Myrs) for a HD~219134 input chemistry and disc model. Figure \ref{fig:HDCT} is a modified version of Figure \ref{fig:HDCD} where we now plot the condensation fraction against temperature rather than radial separation for the two extremes of formation time. Differences for a solar-type input chemistry and disc model are negligible for our purposes (and are therefore not shown).

Figure \ref{fig:HDCD} and Figure \ref{fig:HDCT} illustrate how the model can reproduce the expected condensation series found in \cite{Lodders2003,Lodders2010}. The figures also emphasize how the order of condensation of the elements analysed is invariant over time.


\paragraph{Planetesimal aggregation model}
% introduction to the planet aggregation model
The planetesimal compositions found using the equilibrium chemistry model and the PPD model are a function of formation time and formation location ($t_{\rm disc}, a$). In reality a body the size of a planet will incorporate material from a range of formation locations and possibly a range of formation times and the radial drift of planetesimal could play an important role in the condensates a body aggregates. 
% outline of the simple aggregation model and its inputs
In order to account for these effects, we consider a model in which the material that forms the planets originates from a range of formation locations described by a Gaussian distribution centered at distance $a$ and with a width of $\Delta a$. Thus, we have three free parameters, the formation location, $a$, which is equivalent to the mean of the normal distribution, the feeding zone parameter, $\Delta a$, which is equivalent to the standard deviation of the normal distribution and the formation time of the planetesimals which comprise the planet, $t_{\rm disc}$. At a given formation time, $\Delta a$ corresponds to a temperature range $\Delta T_{\rm cond}$. Figure \ref{fig:sketch} is a schematic diagram highlighting this setup for a given formation time.

\rev{We do not run N-bodys simulations that would allow us to predict the amount of mass available to form a planet. Given our disc model, the available mass in solids between 0.8 and 1.2 AU at 0 Myr is 0.5 \ME. At 2 Myr, the available mass in solids is  even less ($<0.002$ \ME). In order to form a planet on the order of 5 \ME, the disc properties would have to be adjusted, e.g., by making the disc more massive ($M_0 = 0.5 M_*$ instead of $M_0 = 0.1 M_*$) or the surface density gradient steeper ($\Sigma(a,t_{\rm disc}) \propto ({a}/{s_{0}})^{-\nicefrac{37}{19}}$ instead of $\Sigma(a,t_{\rm disc}) \propto ({a}/{s_{0}})^{-\nicefrac{24}{19}}$). Besides some margin for the disc properties, the possible influence of the presence of the outer more massive planets and radial drift of planetesimals from outer disc regions on the available mass in planetesimals in the innermost disc region may be non-negligible. To give an anticipatory example, if planet b were inheriting 1 \ME of material from the innermost and 3.5 \ME from the outer disc, the decrease in bulk density would be limited to 2.4 \%. Here, we assume that sufficient mass is available to form planets of 5 \ME. }

% link to the interiors section
The modelled exoplanetary compositions are used as inputs in the exoplanet interior model outlined in \cite{dorn2017generalized} and the variation in the mass radius curves produced as a function of formation time ($t_{\rm disc}$), formation radius ($a$), and feeding zone size ($\Delta a$) were investigated for the HD219134 system. 

\paragraph{Exoplanet interiors}

The calculated compositions from the condensation model are used as bulk constraints for the rocky interiors of the planets. The employed interior model uses self-consistent thermodynamics and is described in detail by \cite{dorn2015can}.

We assume purely rocky planet interiors that are composed of pure iron cores with silicate mantles. The mantles comprise the oxides Na$_2$O--CaO--FeO--MgO--Al$_2$O$_3$--SiO$_2$ (model chemical system NCFMAS).  Mantle mineralogy is assumed to be dictated by thermodynamic equilibrium and computed by free-energy minimization \citep{connolly2005computation} as a function of composition, interior pressure and temperature. The Gibbs free-energy minimization procedure yields the amounts, mineralogy, and density of the stable minerals.
For the core density profile, we use the equation of state (EoS) fit of solid state iron in the hcp (hexagonal close-packed) structure provided by \citet{bouchet2013ab} on {\it ab initio} molecular dynamics simulations. We assume an adiabatic temperature profile for both mantle and core.
%We assume a solid state iron, since the density increase due to solidification in the Earth's core is small (0.4 g/cm3, or 3\%) \citep{Dziewonski}.

\begin{figure*}
	\includegraphics[width=\linewidth,trim = 2cm 0cm 1cm 0cm,clip]{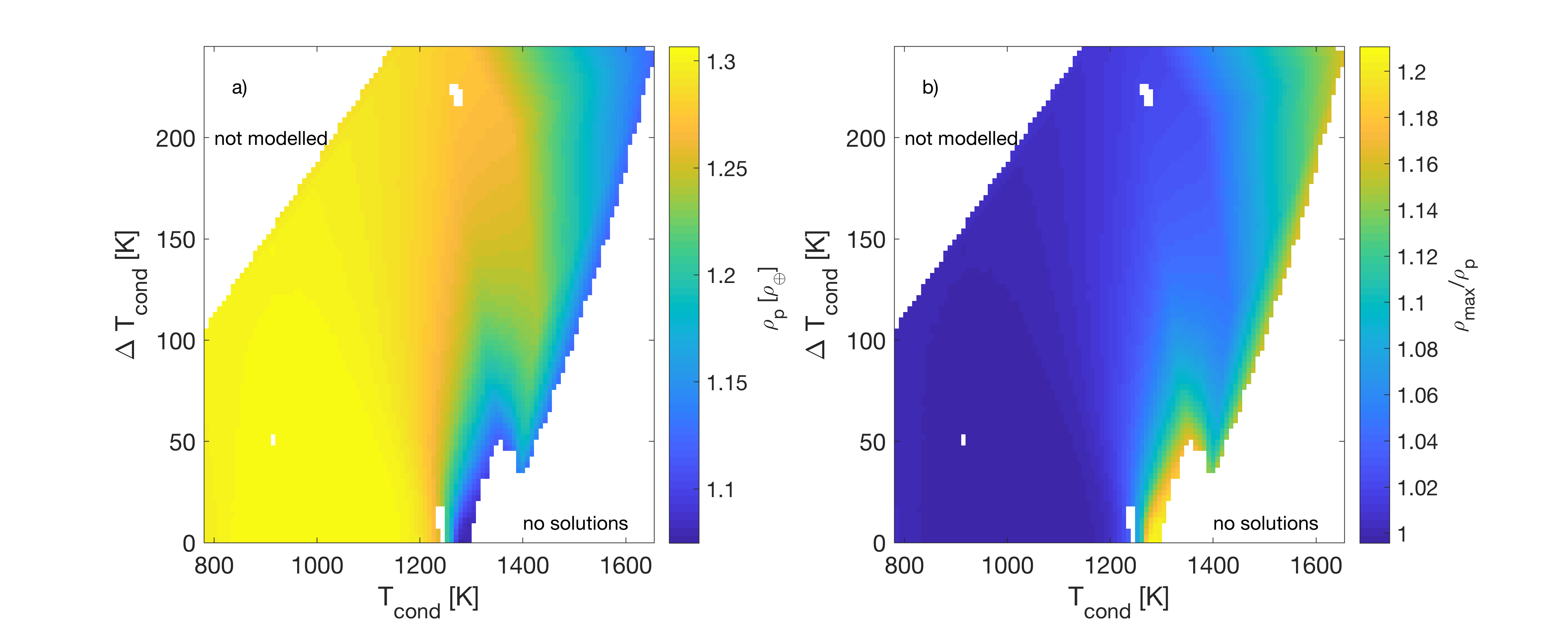}
    \caption{Bulk density of rocky planets built from condensates as a function of the temperature at which the condensates form. In color, (a) planet density $\rho_{\rm p}$ and (b) ratio between the maximum bulk density $\rho_{\rm max}$ and planet density $\rho_{\rm p}$. The maximum bulk density 1.3 \DE is the maximum value of $\rho_{\rm p}$ that is realized at low condensation temperatures (<1200 K). For comparison, densities of planet b and c are $\rho_{\rm b} = 1.15 \pm 0.13$ \DE and $\rho_{\rm c} = 1.26 \pm 0.14 $\DE. }
    \label{fig:rhoT}
\end{figure*}

Given the calculated compositions on Fe, Si, Mg, Al, Ca, and Na (section \ref{sec:HarrisonModel}), we compute the mineralogy and the corresponding bulk density for the given planet masses $M_{\rm c}$ = 4.36 \ME and $M_{\rm b}$ = 4.74 \ME. The compositions vary with the condensation temperatures in the PPD, $T_{\rm cond} \pm \nicefrac{1}{2} \Delta T_{\rm cond}$.  Figure \ref{fig:rhoT} plots the bulk density of planets as a function of condensation temperatures of the corresponding planetesimals. Lowest temperatures are not modelled since changes in bulk densities are expected to be negligible. At high temperatures (>1200 K), planets become very rich in Ca and Al and depleted in Fe. For example, at $T_{\rm cond} =1300$K, the rock composition is made of CaO (16 wt\%), FeO ($\ll0.1$ wt\%), MgO (14 wt\%), Al$_2$O$_3$ (43 wt\%), SiO$_2$ (27 wt\%), and Na$_2$O ($\ll0.1$ wt\%). A planet of this composition is core-free with a stable mineralogy as plotted in Figure \ref{fig:phases} (see Appendix \ref{phases}). For rock compositions where the sum of calcium and aluminium oxides exceed $\sim$80 wt\%, no stable solutions for the mineralogy can be found.

Figure \ref{fig:rhoT}a demonstrates, that compositions dominated by Mg, Si, and Fe corresponding to $T_{\rm cond} < 1200$K explain bulk densities that fit planet c's bulk density $\rho_{\rm c} = 1.26 \pm 0.14 $\DE. The low density of planet b ($\rho_{\rm b} = 1.15 \pm 0.13 $\DE) could be explained with condensates formed at high temperatures, being rich in Ca and Al and depleted in Fe. In that case, planet b has no core. The density ratio $\rho_{\rm max}/\rho_{\rm p}$ is plotted in Figure \ref{fig:rhoT}b and covers the observed value $\rho_{\rm b}/\rho_{\rm c} = 1.1 \pm 0.16$. Thus, the density difference can be related to a difference in rock composition, due to a difference in formation temperature, of the solids out of which planets b and c are built, and hence a difference in their formation location at given times. 

\begin{figure*}
	\includegraphics[width=.92\linewidth,trim = 0cm 0cm 5cm 0cm,clip]{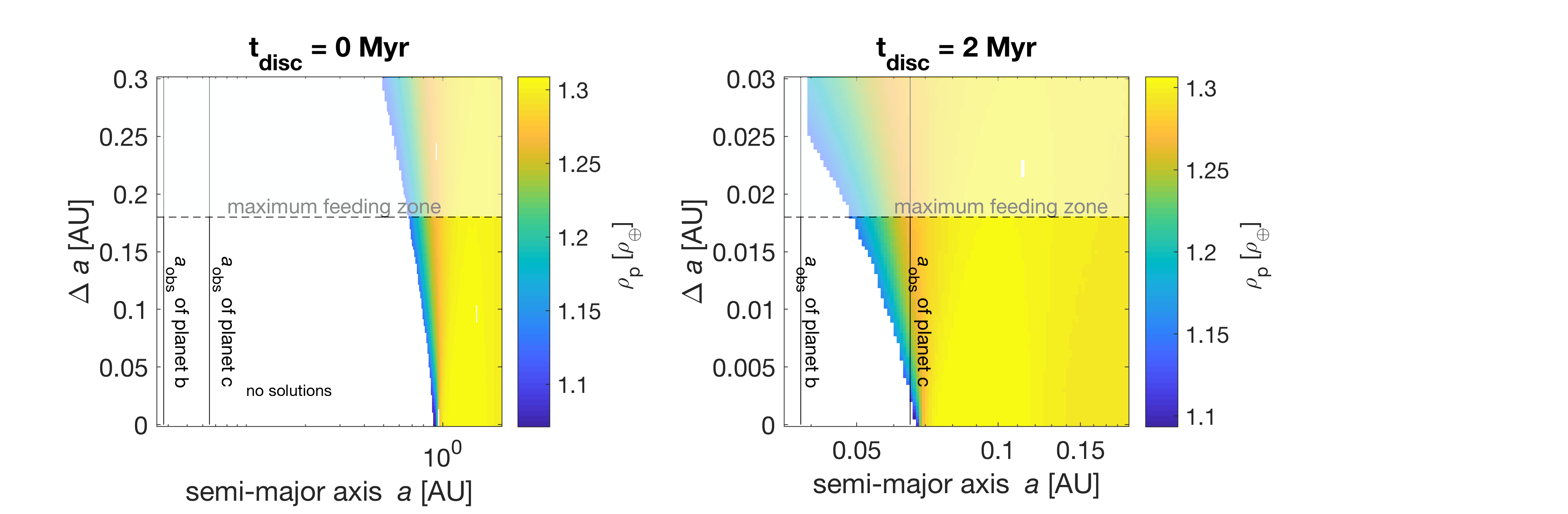}
    \caption{Bulk density of rocky planets built from condensates as a function of the distance to the star \rev{and feeding zone width} at time (a) t$_{\rm disc} = 0$ Myr and (b) t$_{\rm disc} = 2$ Myr. Highest bulk densities are reached further away from the star where PPD temperatures are lower. Rocky planets that are built from high-temperature condensates can have $\sim$15\% lower bulk densities. For comparison, densities of planet b and c are $\rho_{\rm b} = 1.15 \pm 0.13$ \DE and $\rho_{\rm c} = 1.26 \pm 0.14 $\DE. \rev{The maximum feeding zone width corresponds to 10 Hill radii}.}
    \label{fig:rhoAU}
\end{figure*}

We plot the bulk densities of planets as a function of $a \pm \nicefrac{1}{2} \Delta a$ for $t_{\rm disc} = 0$\,Myrs and $t_{\rm disc} = 2$\,Myrs in Figure \ref{fig:rhoAU}. \rev{The feeding zone $\Delta a$ is generally mass dependent in oligarchic growth and is usually set to a maximum of 10 Hill radii \citep{ida2004toward}. For planet b at 1 AU ($t_{\rm disc} = 0$\,Myrs) this maximum equals 0.18 AU, while at 0.1 AU ($t_{\rm disc} = 2$\,Myrs) it is 0.018 AU (Figure \ref{fig:rhoAU}). Larger effective feeding zones may be realized as a result of scattering of planetesimals by neighboring planets.}

\rev{In Figure \ref{fig:rhoAU}, we show two different times during the disc evolution, 0 and 2 Myr. Two million years is a general time span for which gas in the PPD can be present. However, from recent disc surveys it seems that the majority of planetesimal formation occurs very early for the discs investigated here and maybe limited to $<1$ Myr \citep{tychoniec2018vla}.} During this time, planets can exchange angular momentum with the disc gas and migrate. When the gas has dissipated, migration of planets ceases. Given our model assumptions, the orbital evolution of the planets can be informed by their bulk densities. 
With respect to Figure \ref{fig:rhoAU}, planet c could have started forming at $a \geq 0.9 AU$, migrating within 2\,Myrs to its current position $a_{\rm obs,c} = 0.0653 AU$. Planet b could have started its formation at $ 0.5 < a \leq 0.9 AU$ and migrated to its current position within 2 Myr.

In summary, the density differences of both Super-Earths can be inherited from the chemical variability of rocky planetesimals due to temperature differences in the PPD. Hence, the density differences of the planets constrain the temperatures of accreted building blocks and thus relate to formation locations and  migration histories. In the following we consider whether the inferred formation locations and present observed locations are compatible with a model in which the two planets undergo simultaneous, disc-driven migration.

\subsubsection{Migration pathways}
\label{formation}
Given the close proximity of the planetary system - particularly planet b - to the host star, it seems likely that some amount of migration was necessary to assemble the final system. The rate of type I migration is linearly proportional to the mass of the planet \citep[see e.g][]{Baruteau2014}, meaning that one would naively expect planet b to migrate around 9\% faster than planet c, given its marginally higher mass. This raises the possibility that planet c originally formed interior to planet b, but was caught and eventually overtaken by its more massive sibling during migration. In order for this to occur, planet b would have to migrate significantly faster than planet c, such that the two planets would not trap in a first order mean-motion resonance which would keep them well separated and stop them from switching positions. Given the 9\% mass difference, such a sizeable difference in migration time-scale seems unlikely. To investigate this possibility, we followed the method used in \cite{Hands2014}, using $N$-body models of planets b and c with additional parametrised forces to represent the interaction between the two planets and the disc. These parametrised forces enforce migration as follows
\begin{equation}\label{eqn:mig}
\frac{da}{dt} =\frac{a}{\tau(M_p)}
\end{equation}
where $a$ is semi-major axis, $\tau(M_p) = \tau_0 M_c/M_p$, $M_c$ is the mass of planet c and $M_p$ is the mass of a given planet undergoing migration. There is also a dimensionless eccentricity damping parameter $K$ that defines the speed of eccentricity damping relative to the migration times-scale (larger $K$ results in faster damping). We started each simulation with planet c at 1 AU and planet b at 1.5 AU, putting the two planets exterior to the 3:2 mean-motion resonance, and stopping each simulation when either planet reached the observed location of planet b. We tried $\tau_0 = $100, 1000, 10,000 and 100,000 yr, with no eccentricity damping and $K=100$. No combination of these parameters allowed planet b to catch and overtake planet c. Indeed, the two planets trapped in every case in either the 3:2 or 4:3 mean-motion resonance. We experimented with adding stochastic forcing to represent the force of a turbulent disc on the two planets \citep{Rein2009}, but were only able to exchange the positions of the two planets with extremely high levels of stochastic forcing.

Due to the apparent difficulty switching the positions of these two planets, we suggest that they most likely formed in their present order and then migrated together into the inner disc. Assuming planet b formed at 1 AU and migrated at a rate 9\% faster than planet c, one can show using equation \ref{eqn:mig} that planet c would need to have formed at 1.3 AU to explain the relative locations of the planets today. This difference in formation location is in agreement with the argument presented in section \ref{Sec:refr} that planet b may be formed from high temperature condensates while planet c is built from condensates below 1200 K. We confirmed using the same parametrised $N$-body method that the two planets do not trap in resonance if subjected to divergent migration in this manner.

There are of course additional complications to this picture. \rev{The other inner planets f and d, exterior to b and c,} are significantly more massive than b and c based on the lower limits of their masses. If planet f and d also migrated in the type I regime, they would certainly migrate faster than and catch the inner two planets, trapping them in resonance. However, the much larger masses of the outer planets could cause them to open a gap in some regions of the disc, leading to slower migration in the type II regime. \rev{We calculate a gap-opening criterion $\mathcal{P}$ \citep[see equation 9,][]{Baruteau2014} and assume a moderately flaring disc with scale-height $H = 0.033 R^{5/4}$ and a \cite{Shakura1973} viscosity parameter $\alpha = 1 \times 10^{-2}$ to understand if this fate might befall the planets. Indeed, with this disc model and using the lower limits on the masses of planets \rev{f and d, all four inner planets} would be marginally gap opening at their present locations, with $\mathcal{P}$ between 4 and 7. \cite{Crida2007} show that planets with these values of $\mathcal{P}$ open gaps of between 50 and 70\%, and that for such partial gaps, material left in the partial gap can exert a positive torque on the planets. This extra torque can cause migration to proceed more slowly than standard type II, or even drive outward migration.}

Of course these results are dependent upon the disc model - thicker and more viscid discs would prevent all four inner planets from opening any gap, while a higher flaring index would allow them to open gaps further out in the disc. In any case though, if the scale-height of the disc does flare with radius, the more massive outer two planets \rev{f and d} would open gaps further out in the disc than their lower-mass counterparts and therefore their migration would slow first. The depth of the gaps and their formation radius depends upon the exact masses of the planets \rev{f and d}, and thus if they are indeed more massive than the lower limits, then they could potentially have opened gaps even further out in the disc. We suggest then, that all four \rev{inner} planets formed in their present order, and that planets f and d initially migrated in the type I regime and moved closer to planets b and c until they reached a region of the disc where they could open partial gaps. At this point they entered the modified type II regime - where coorbital material reduces the type II migration rate - and planets b and c were able to continue their migration unhindered. We note however that the migration of four planets in unison is a complicated and non-linear problem that is highly dependent on disc parameters, and further work - including hydrodynamical simulations - is required to understand the most likely evolution pathways of these inner four planets in their host disc. For instance, in the partial gap opening regime there is a potential for migration to runaway, leading to very fast inward migration \citep[see e.g.,][]{Baruteau2014}. \rev{Given their enormous orbital separation relative to the rest of the system, ee do not expect the outermost planets g and h \citep{motalebi2015} to influence this picture.}
We further note that recent work \citep[e.g.,][]{Goldreich2014,Hands2018} suggests that mean-motion resonances created during migration might also be broken by the disc, and therefore the lack of resonances in the present-day system does not necessarily mean that the system was always free of them.

% \begin{table}
% 	\centering
% 	\caption{This is an example table. Captions appear above each table.
% 	Remember to define the quantities, symbols and units used.}
% 	\label{tab:example_table}
% 	\begin{tabular}{lccr} % four columns, alignment for each
% 		\hline
% 		A & B & C & D\\
% 		\hline
% 		1 & 2 & 3 & 4\\
% 		2 & 4 & 6 & 8\\
% 		3 & 5 & 7 & 9\\
% 		\hline
% 	\end{tabular}
% \end{table}
\subsection{Different volatile budgets as a cause for lower bulk density of planet b}
\label{Sec:vol}
In this Section, we discuss and investigate whether the density difference between planet b and c can be explained solely by different volatile layers while neglecting any differences in rock composition as discussed before (Section \ref{Sec:refr}). In general, possible volatile layers include (1) primordial atmospheres, (2) outgassed atmospheres, (3) water-rich layers. We discuss each of the three possibilities and attempt to quantify their probabilities.

\subsubsection{Primordial atmosphere}
\label{sec:primordial}
Hydrogen-dominated primordial atmospheres for both planets have been excluded by \citet{dorn2018secondary}. They suggest a theoretical minimum threshold-thickness for a primordial atmosphere below which evaporative loss efficiently erodes an atmosphere on short time-scales (tens of Myr). The minimum threshold-thickness for H-dominated atmospheres on planet b and c are significantly larger than the inferred atmospheric thicknesses implying that the atmospheres must be of higher mean-molecular weight and/or is outgassed from the rocky interior.

\begin{figure}
	\includegraphics[width=0.95\columnwidth,trim = .1cm 0cm 0.3cm 0cm,clip]{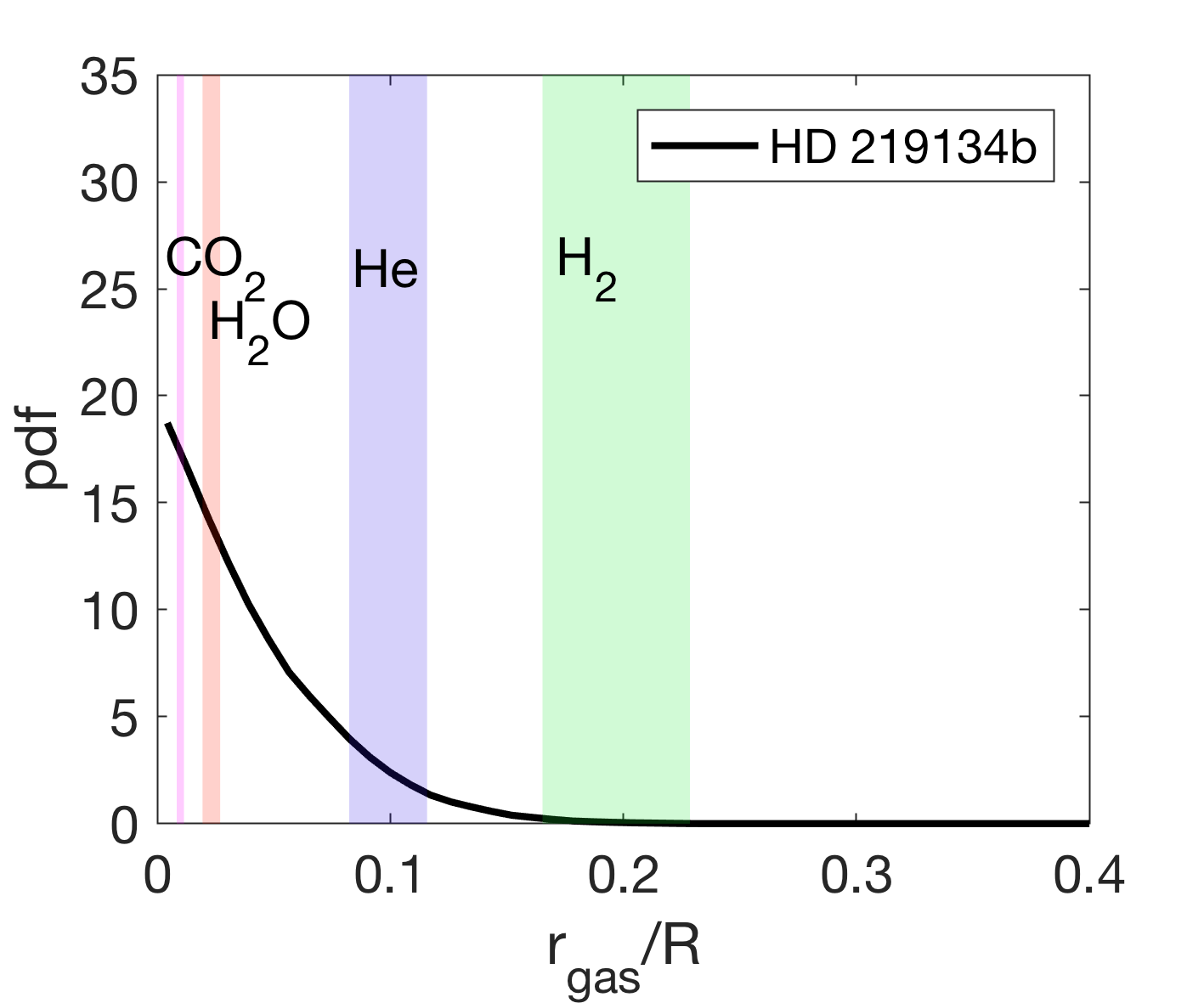}
    \caption{Comparison between inferred gas layer thickness (black curves) for planets b and the minimum threshold thicknesses $\Delta R_{\rm th}/R$ for gas layer dominated by H$_2$ (green), He (blue), H$_2$O (red), or CO$_2$ (magenta). The \rev{probability density function (pdf) of} inferred gas layer thicknesses (black line) are taken from \citep{dorn2018secondary}. If the inferred gas layer thicknesses were significantly smaller than a given $\Delta R_{\rm th}/R$, the corresponding species would be excluded to dominate the gas layer. For both planets, the gas is very unlikely to be dominated by H$_2$ and He to some extent.
     Minimum threshold thickness are estimated by Equation \ref{Eq:threshold}. The uncertainties in $\Delta R_{\rm th}/R$ include the uncertainties in planetary mass and radius, and evaporative efficiency ($0.01<\eta<0.2$).}
    \label{fig:Rthres}
\end{figure}

While \citet{dorn2018secondary} focused on H-dominated atmospheres, we expand their methodology to also calculate the minimum threshold-thicknesses for atmospheres dominated by He, H$_2$O, and CO$_2$. Figure \ref{fig:Rthres} shows the comparison of the different threshold-thicknesses with the inferred atmospheric thicknesses as taken from \citep{dorn2018secondary}. \rev{The probability distribution on atmospheric thickness was calculated using a Bayesian inference analysis using the data of planetary mass and radius, and relative refractory element abundances based on the host star proxy. Their interior model allowed for variability in core size, mantle thickness and composition, water mass fraction, atmosphere composition and extent, and thermal state. We employ the same model for the rocky interior.}

The minimum threshold-thicknesses $\Delta R$ are calculated following the methodology of \citet{dorn2018secondary} but with one adaption: The minimum threshold-thicknesses corresponds to a minimum mass of gas. While \citet{dorn2018secondary} estimate this gas mass by the amount of gas that is lost during the age of the star, we use the amount of gas that is lost over 100 Myr. Using stellar age is suitable for primary or primordial atmospheres only. Differences between our results and \citet{dorn2018secondary} are small. Details on the calculation of the minimum threshold-thicknesses $\Delta R$ and mass loss rates $\dot{m}_{\rm gas}$ are discussed in the Appendix \ref{App:massloss}.

Among all thresholds for H$_2$, He, H$_2$O, and CO$_2$ dominated atmosphere, only the threshold $\Delta R$ for H$_2$ (and to some extent He) is larger than inferred gas layer thicknesses (Figure \ref{fig:Rthres}). Thus, as found by \citet{dorn2018secondary}, a primordial H-atmosphere can be excluded and a helium-atmosphere is little likely for planet b.

% gas layer could be high metallicity-type, outgassed from magma-ocean or over long-term

\subsubsection{Outgassed atmospheres}

Generally, outgassed atmospheres of higher mean molecular weight (e.g., H$_2$O, CO$_2$, CO, CH$_4$) can originate from early outgassing during the magma ocean stage (primary atmosphere) or during solid state evolution of a planet (secondary atmosphere). Could these also explain the lower density of planet b? 

\begin{figure}
	\includegraphics[width=\columnwidth]{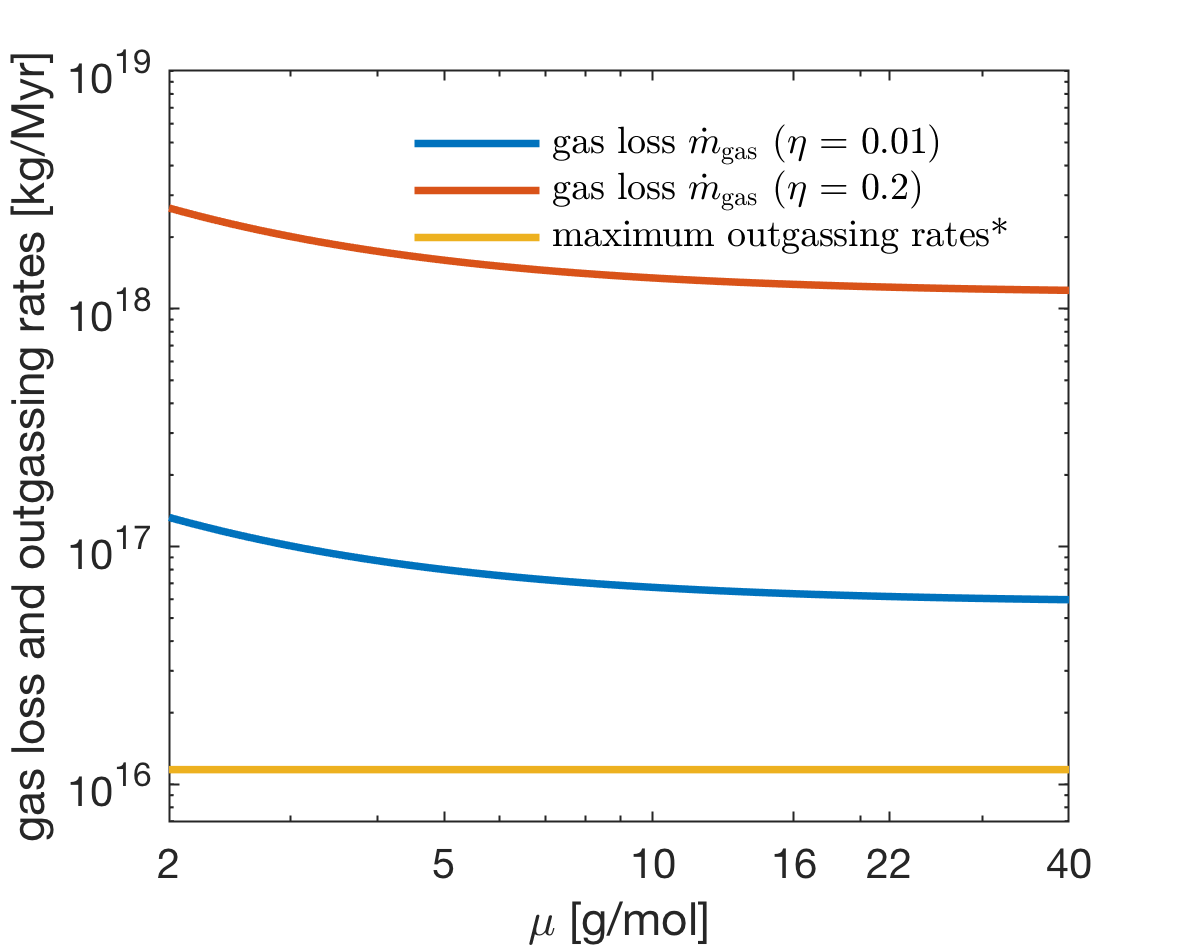}
    \caption{Planet b: Gas loss and gain rates in comparison as a function of gas mean molecular weight. Gas loss rates are estimated by Equation \ref{eq:masslossrate} for two different evaporative efficiencies (blue and red curves). Gas gain rates are based on possible outgassing from the rocky interior for which we neglect dependencies of gas composition. The shown maximum outgassing rate is based on \citet{kite2009geodynamics} for planet masses of 4-5 \ME and ages 8-13 Gyrs and is in agreement with \citet{noack2014can,tosi2017habitability,dorn2018outgassing}. The outgassing rates are lower compared to the loss rates, suggesting that a build-up of a secondary gas layer on planet b over geological time is difficult. }
    \label{fig:massloss}
\end{figure}

Secondary atmospheres outgassed during the solid state evolution of planets continuously replenish the atmosphere over geological time-scale. The rate of outgassing depends on numerous aspects of which planet mass, planet age, thermal state, and convection mode (e.g., stagnant-lid, mobile-lid) dominate. The latter two are largely unconstrained \citep[e.g.,][]{valencia2007inevitability,van2011plate,tackley2013mantle,oneill2014mantle,noack2014plate}. However, planets much older and 4-5 times larger than Earth are expected to experience less outgassing than Earth-like planets.
Given estimates of planet mass and age (12.9 Gyr as estimated by \citet{takeda2007structure}), an upper limit on the outgassing rate can be stated. In Figure \ref{fig:massloss}, we show an upper outgassing rate as extracted from \citet{kite2009geodynamics} for planet masses of 4-5 \ME and ages of 8-13 Gyr. This maximum outgassing rate neglects chemical processes and is thus independent of $\mu$; it includes the uncertainty on the convection mode and is comparable to other studies \citep{dorn2018outgassing,tosi2017habitability}. 

The upper outgassing rate is a factor of few lower than our estimates of evaporative loss rates (Eq. \ref{eq:masslossrate}). Although uncertainties on both loss and outgassing rates may be significant \citep[e.g.,][]{tackley2013mantle,noack2012coupling,tosi2017habitability,lopez2017born,erkaev2013xuv}, volcanic replenishment of the planet b's atmosphere seems limited given its continuous erosion by stellar irradiation.

Can a primary atmosphere that is outgassed from the early magma ocean stage be present on planet b and explain its low density? Let us assume a maximum volatile content in silicates of 3\% that is the maximum water content measured in achondritic meteorites \citep{jarosewich1990chemical} that are discussed as proxies for silicate parts of differentiated planetesimals \citep{elkins2008ranges}. Multiplying 3\% with the planet's mantle mass, this translates into a maximum volatile content of $4 \times 10^{23}$ kg (assuming an Earth-like 50 \% core mass fraction). During the age of the planet (12.9 Gyr), about $1 \times 10^{23}$ kg of volatiles are lost (for low evaporation efficiencies of $\eta = 0.01$ and assuming Sun-like evolution the X-ray flux as described by \citet{ribas2005evolution} \footnote{The X-ray flux evolution of the Sun is $F_{\rm X} \propto t_{\rm Sun}^{-1.83}$ for $t_{\rm star} > 0.1 {\rm Gyr}$, and $F_{\rm X}$ at 0.1 Gyr is the maximum value of $F_{\rm X}$ for $t < 0.1$ Gyr (Ribas et al., 2005)}). Thus the maximum amount of gas that can be outgassed during the magma ocean stage is of  similar order than the minimum amounts that are lost by stellar irradiation. This simple back-of-the-envelope calculation shows that the amount of primary atmosphere on planet b must be limited.
However, most importantly, there is no reason to believe that planet b could have a primary atmosphere, while planet c could not since both planets are similar in mass.
%, but planet b is the innermost of the two, the difference between the planets that we expect is that planet b would be built from material that is poorer in volatiles compared to planet c. This statement of course assumes that both planets never exchanged their positions, which we discuss in Section \ref{sec:waterlayer}.
Also, the closer distance to the star of planet b implies that planet b would lose 3 times more gas mass than planet c. Thus, a massive primary atmosphere on planet b is unlikely to explain a density difference between the planets of 10\%.

\begin{figure}
	\includegraphics[width=\columnwidth]{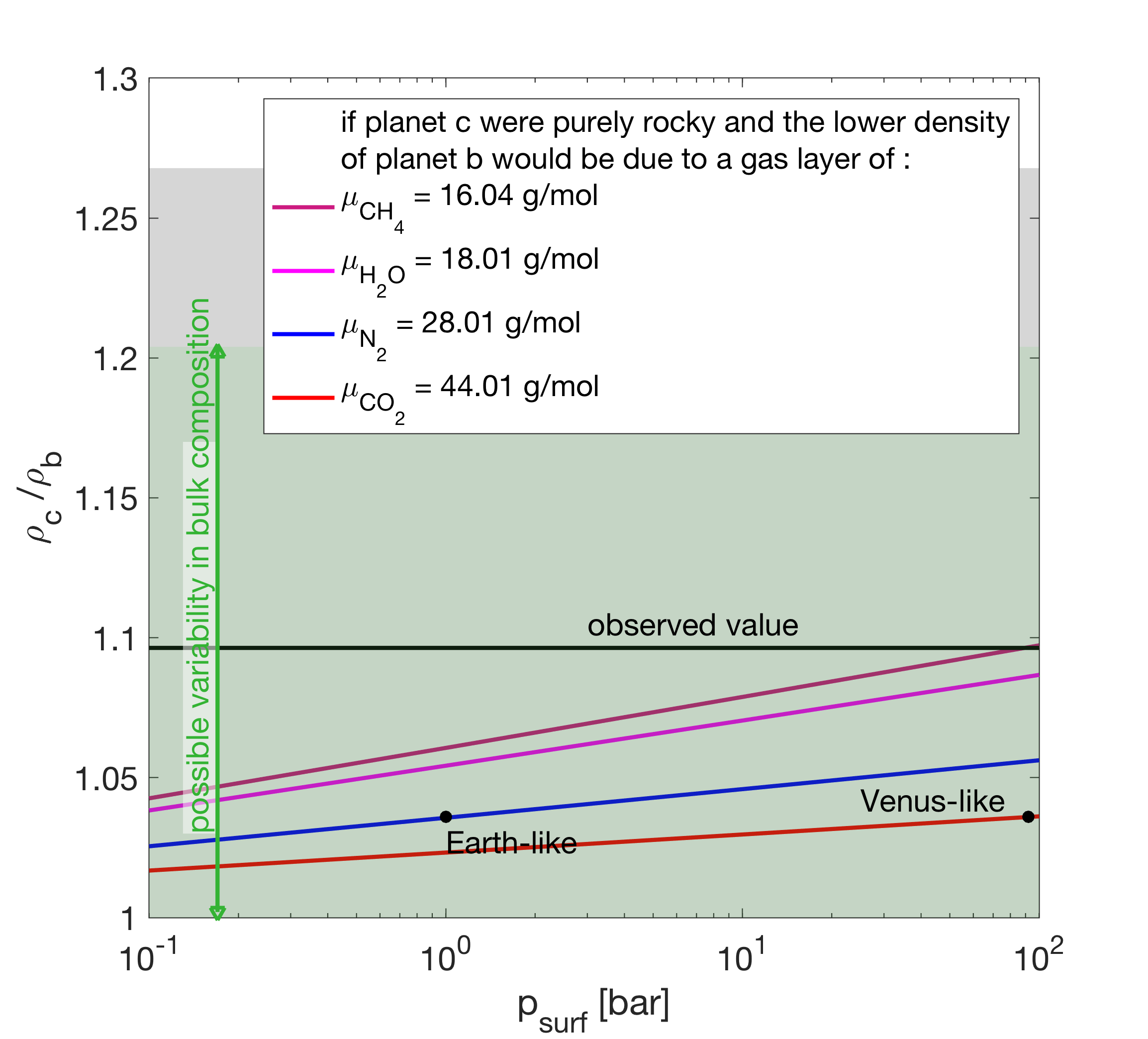}
    \caption{Possible scenarios for the different bulk densities of planet b and c. The observed value ($\rho_{\rm c}/\rho_{\rm b} = 1.10 \pm 0.16$) can be explained by three scenarios: (1) within observational uncertainties, planet b and c can have the same bulk density and thus similar interiors, (2) planet b is lower in bulk density due to a outgassed gas layer (colored curves for different gas species), (3) planet b is lower in density due to possible compositional variability of rocky building blocks as a consequence of temperature variations in the PPD (green area). The latter scenario can explain 85\% of the observed value, while a outgassed gas layer can explain 35\% (for Venus- and Earth-like atmosphere compositions of up to 100 bar) to 50\% (for CH$_4$ atmospheres up to 100 bars) of the observed value.  }
    \label{fig:gas}
\end{figure}

In summary, it is unlikely that a massive primary or secondary atmosphere is present on planet b but not on planet c. Our above argumentation is rather qualitative, a quantitative description would involve many unconstrained interior parameters and lies outside the scope of our paper. However, we quantitatively investigate the hypothetical scenario of planet b hosting a terrestrial-type atmosphere while planet c is a bare rocky planet. This extreme scenario maximizes the modelled density difference between both planets. 
Figure \ref{fig:gas} plots the bulk density ratio of planet c to planet b ($\rho_c/\rho_b$). The observed value is 1.1 with large uncertainties (grey area). If planet c were purely rocky and planet b were simply a scaled version thereof but also including a terrestrial-like atmosphere, the density ratio can reach values as high as 1.035 (for Earth- or Venus-like atmospheres). Larger density ratios can be reached with atmospheres of larger scale height that are dominated by gas of lighter mean molecular weights (e.g., H$_2$O, CH$_4$). A 100 bar methane atmosphere on planet b can indeed explain $\rho_c/\rho_b = 1.1$.
Allowing for differently thick ($\leq 100$ bars) terrestrial-like atmospheres (N$_2$, CO$_2$) on planet b, 35\% of the observed density data can be explained, while lighter mean molecular weight atmospheres (H$_2$O, CH$_4$) can explain 50\% of the observed densities. The probabilities (35\% and 50 \%) are simply derived by integrating the normal distribution of $\rho_c/\rho_b = 1.1 \pm 0.16$ from zero to 1.035 (obtaining 35\%) or 1.1 (obtaining 50\%), respectively.

% gas layer replenishment over long-term geological cycles is hard because outgassing is limited at this age and mass

% possible gas layers can explain the difference of bulk densities to xx%

\subsubsection{Water-rich layers}
\label{sec:waterlayer}

The low density of planet b can in principle be explained by water layers of few percents in mass fraction. This scenario is only possible, if planet b formed outside of planet c and both planets exchanged their positions after their formation. In Section \ref{formation}, we have shown that the repositioning of the planets during migration is unlikely. % given their similar migration time-scales and the observed system architecture.

%%%%%%%%%%%%%%%%%%%%%%%%%%%%%%%%%%%%%%%%%%%%%%%%%%%%%%%%%%%%
%%%%%%%%%%%%%%%%%%%%%%%%%%%%%%%%%%%%%%%%%%%%%%%%%%%%%%%%%%%%
%%%%%%%%%%%%%%%%%%%%%%%%%%%%%%%%%%%%%%%%%%%%%%%%%%%%%%%%%%%%

\subsection{Observational biases}
\label{Sec:bias}

Observational biases on planetary data are generally difficult to quantify. They can originate from limited amount of observations and inaccurate model assumptions. Two main sources are inaccuracies on stellar and orbital parameters. A difference in stellar parameters would equally effect data of both planets b and c, and thus the \emph{relative} difference in their bulk densities - which is the focus of this study - would be largely unaffected. Also, this nearby star HD219134 is very well characterized. Its density was derived by using stellar models and was further validated by \citet{gillon2017two} using a global analysis of the transit photometry.
Unresolved orbital parameters can influence the determined planet mass. For example, a circular orbit was assumed for planet b given its estimated short circulation time-scale of 80 Myr \citep{gillon2017two}. Assuming the hypothetical case that planet b could have an eccentricity similar to planet c, the resulting mass and bulk density of planet b would be even slightly smaller (by 0.002\%), which would support an existing difference in bulk densities between the planets.

Further sources of bias are instrumental systematics. For the transit photometry, the employed 4.5 $\mu m$ detector of Spitzer/IRAC \citep{gillon2017two} has an accuracy on the scale of photon-noise and thus possible bias is within one standard deviation \citep{ingalls2016repeatability}. Accuracy on the HARPS-N RV dataset is high thanks to an extremely stable and strictly
controlled instrument and a tailored data reduction software. A total instrumental error of $\sim$0.5 m/s is expected \citep{cosentino2014harps} which is large compared to the measured RV signal error of 0.075 m/s. Dominating instrumental errors stem from spectral drifts due to temperature and air pressure variations at the HARPS-N site. Since the RV measurements for both planets were taken 
from the same time period (August 2012 - September 2015), we expect the relative masses ($M_{\rm b} - M_{\rm c}$) to be more accurate than the individual masses ($M_{\rm b}$, $M_{\rm c}$). However, it lies outside of the scope of this study to quantify the possible accuracy on the relative masses. %The limited accuracy of the planetary masses is probably one critical aspect for our study, since it is the small bulk density difference of 10 \% which we investigate here.

For the nominal data uncertainties (Table \ref{tab:planetdata}), there is a 20\% chance that both planets have a similar bulk density (within 5\%, like Venus and Earth). If we add the instrumental error of 0.5 m/s to the nominal RV signal, the uncertainties on the bulk densities become large (b: 24\%, c: 34\%). In that case, the chance that both planets have similar bulk densities are lower (8\%), which is simply because larger data uncertainties imply a larger range of possible densities for each planet.

\subsection{Discussion}
\label{discussion}

Here, we have argued that the density difference between planet b and c is likely due to a difference in rock composition instead of a difference in volatile layer thickness. While planet c can be explained by silicates and iron (dominantly Mg, Si, Fe, O), planet b can be dominated by a Ca and Al-rich interior with no iron core. This drastic difference in rock composition has implications on their possible interior dynamics, magnetic fields, and atmospheric properties.

%further constraints and expectations on their atmospheres
The star is bright and small enough to allow the characterization of the planetary atmosphere in terms of composition and extent. Follow-up transit transmission spectroscopy with the Hubble Space Telescope (HST) and occultation emission spectroscopy with James Webb Space Telescope (JWST) were suggested by \citet{gillon2017two}. On one hand, the addition of constraints on atmospheric composition and extent may provide valuable information to further judge whether planet b's low density is indeed not due to a thick atmosphere but due to an Ca and Al-rich interior. 
On the other hand, if possible atmospheres on both planets interact with their rocky interiors, we naively expect that possible terrestrial-type atmospheres on both planets may be chemically very different, e.g., due to different oxidizing conditions \citep{palme1990high}. However, both planets b and c can have no significant atmosphere. 

Furthermore, the Ca and Al-rich interior of planet b implies the absence of an iron core. Thus, \rev{a magnetic field as generated by core dynamics like on Earth cannot exist for} planet b. If interactions between the magnetic fields of star and planet become detectable \citep{saur2013magnetic}, \rev{this system could be an interesting target to investigate if signatures for planet b and c differ significantly}.

% improved masses and radii
Further observations will also allow us to improve the precision on planet radii and masses as discussed by \citet{gillon2017two}. With at least 50 observed transits for both planets and at least 2000 new RV measurements, the precision on planet radii can be  less than 1 \% and 3\% on planet mass. This would lead to a precision of 4\% on bulk density and thus 6\% on the density ratio $\rho_{\rm c}/\rho_{\rm b}$. Within 1-$\sigma$, this improved uncertainty could rule out whether both planets are scaled-up analogues with no considerable difference in their bulk densities.

\begin{figure}
	\includegraphics[width=\columnwidth]{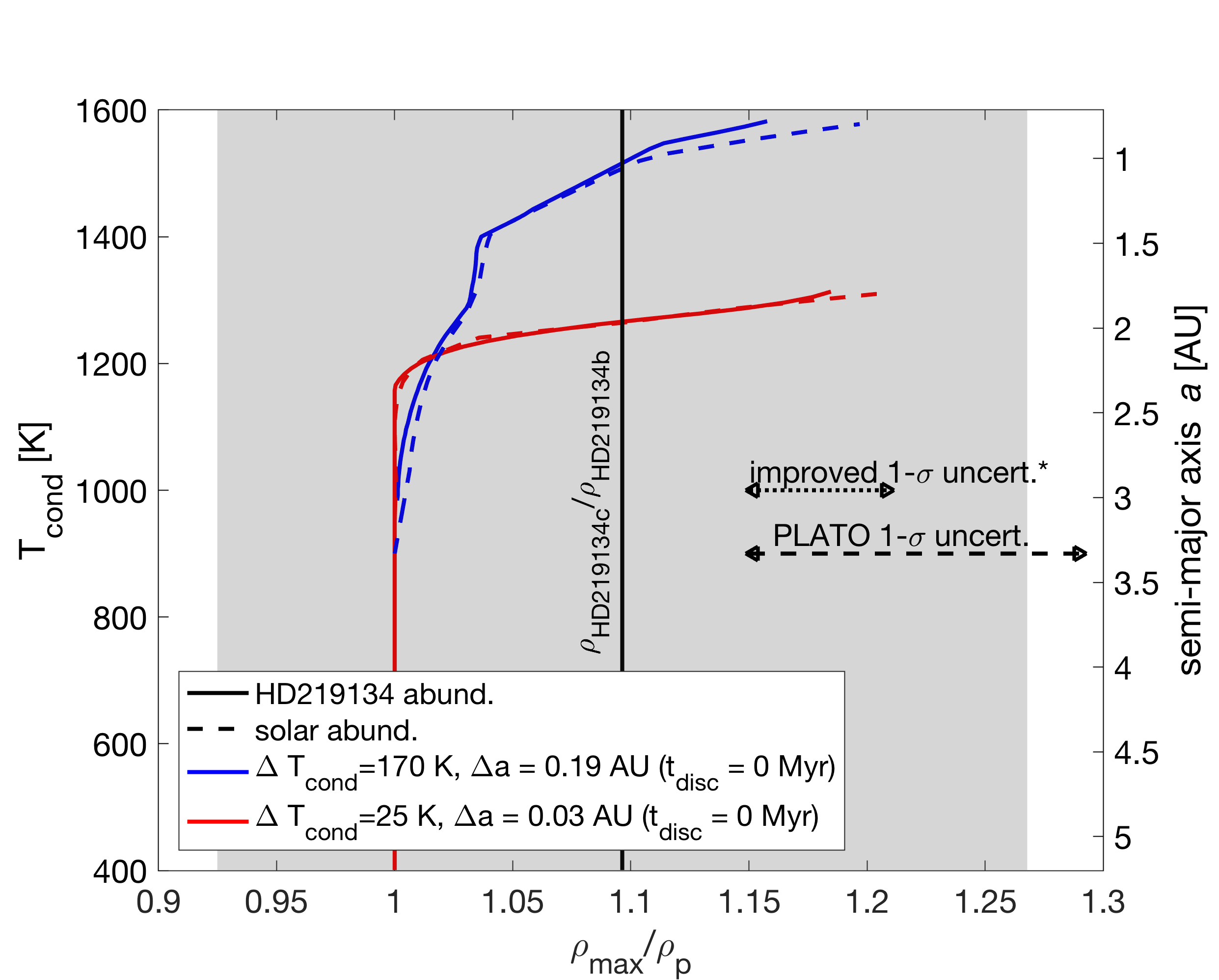}
    \caption{Possible variability in bulk density for a rocky planet plotted as the ratio $\rho_{\rm max}/\rho_{\rm p}$ as a function of temperature T$_{\rm cond}$ \rev{and semi-major axis $a$ (at $t_{\rm disc} = 0$\,Myr)}. If a planet was built from condensates that formed at temperatures T$_{\rm cond}$, its density $\rho_{\rm p}$ could considerably deviate from the $\rho_{\rm max}$ that is realized for the majority of planets which form at temperate conditions (<1200 K). $\Delta T_{\rm cond}$ is evaluated at 0 Myr. The grey bar denotes 1-$\sigma$ uncertainty of $\rho_{\rm HD219134c}/\rho_{\rm HD219134b}$. $^*$Further observations of HD219134 (see Section \ref{discussion}) can yield data precision that allows better constraints on the potential of planet b being built from high temperature condensates (4\% precision on bulk density). Also, the expected 1-$\sigma$ uncertainty for PLATO data is shown (10\% precision on bulk density). }
    \label{fig:plato}
\end{figure}

\subsection{Summary on HD219134 b and c}

For HD219134, the innermost planet b is 10 \% lower in bulk density compared to planet c. We investigated how the interiors could differ from each other in order to reproduce the observed densities. The probabilities for a difference in volatile layers is 35-50 \%, while both planets can also be similar (like Earth and Venus) with a probability of 20 \%.
The most likely scenario (80\%) is a difference in rock composition, i.e., planet b can be explained by a Ca and Al-rich composition that is inherited from the first solids to condense out of the disc (above 1200 K), while planet c's interior is best fit with a composition rich in Mg, Si, and Fe as we know it from Earth. With a relatively simple migration model, we show that  corresponding formation histories of both planets are plausible. Assuming planet b started its formation at 1 AU, planet c would need to form at 1.3 AU such that both planets reach their observed positions during the gas disc life.

\section{55~Cnc~e}
\label{Sec:55cnc}
\subsection{Previous studies on 55~Cnc~e interior}

There are plenty of studies discussing the possible nature of the highly irradiated 55 Cnc e \citep[e.g.,][]{gillon2012improved,demory2016map,bourrier201855}. \citet{dorn2017bayesian} have first used the stellar abundances as proxies for its rocky interior. Its latest mass-radius estimates (8.59 \ME, 1.95 \RE) \citep{crida2018update} imply non-negligibly thick layers of volatiles under the assumption that the lowest rock density is the one of MgSiO$_3$.  Different possible volatile layers were proposed. Water envelopes with fractions of some percents were investigated by \citet{lopez2017born, bourrier201855}. Hydrogen layers were discussed by \citet{gillon2012improved,hammond2017linking}. However, the presence of both water or hydrogen are difficult to explain given the non-detection of hydrogen in the exosphere \citep{ehrenreich2012hint, esteves2017search}. \rev{Also, a 10 bar hydrogen layer is implausible given its short life time of less then 1 Myr \citep{lammer2013probing}}. High metallicity envelopes are analyzed by \citet{demory2016map,crida2018mass, bourrier201855} but their inferred gas mass fraction are difficult to achieve from interior outgassing.% \citep{kite2009geodynamics}.

The intense stellar irradiation leads to escape of gas, which has been characterized to \rev{possibly} contain ionic calcium and sodium (during a single epoch only)  \citep{ridden2016search}. A mineral-rich and water-depleted atmosphere was indeed predicted by \citet{ito2015theoretical} assuming that gas and possible molten surface rocks are in equilibrium. Such atmospheres can only reach surface pressures of $\le$0.1 bar. Qualitatively, a mineral-rich atmosphere could explain the relatively high night-side temperature \citep{zhang2017effects}.

Later studies favor a rocky planet with a thin mineral-rich atmosphere \citep{demory2015variability, crida2018mass, bourrier201855}. However, within the data mass-radius uncertainty, interiors with plausible ranges of mineral-rich  atmospheres ($\le$0.1 bar surface pressures) can only explain high density interiors but fail to explain the low density interiors of the mass-radius distribution \citep{bourrier201855}.

\subsection{A possible Ca-Al-rich interior}

\begin{figure}
	\includegraphics[width=1\columnwidth,trim = .2cm 0cm 0.8cm 0.1cm,clip]{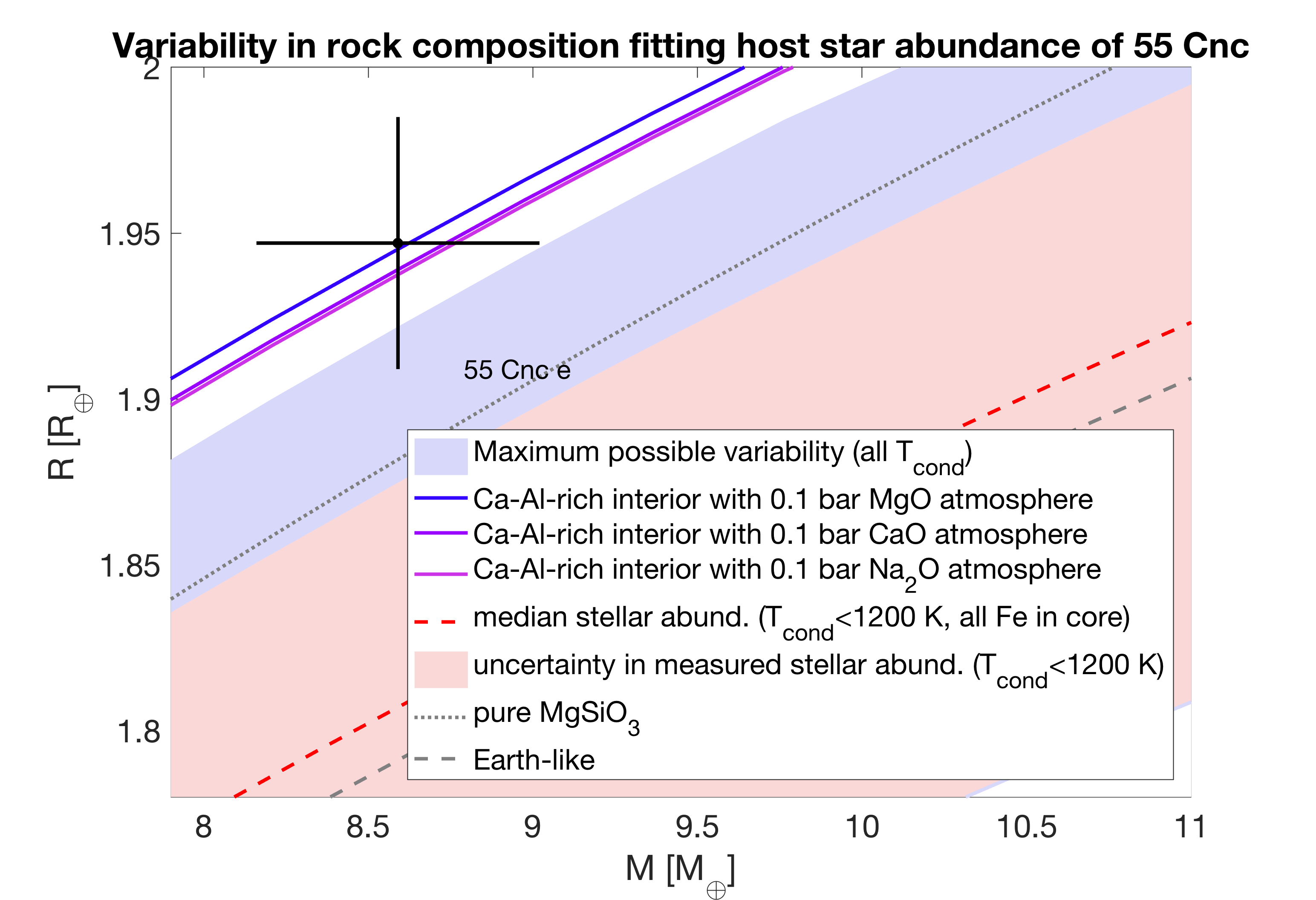}
    \caption{Mass-radius plot showing 55 Cnc e compared to scaled idealized interiors. We assume a 1:1 compositional relationship (i.e., ratios of refractory elements) between the host star 55 Cnc and the PPD in which 55 Cnc e formed. The variability of purely rocky planets forming at different times and locations within the disc is highlighted by the blue area.
    Purely rocky interiors that are built from temperate condensates ($T_{\rm cond}<1200$K) follow the red area, which respects the uncertainty in measured stellar abundances. Interiors that fit the median stellar abundance follow the red curves. The solid colored curves assume the lowest density of a Ca and Al-rich composition plus a 0.1 bar mineral atmospheres.
    }
    \label{fig:MR_55cnce}
\end{figure}

\begin{figure}
	\includegraphics[width=\columnwidth,trim = .1cm 0cm 0cm 1cm,clip]{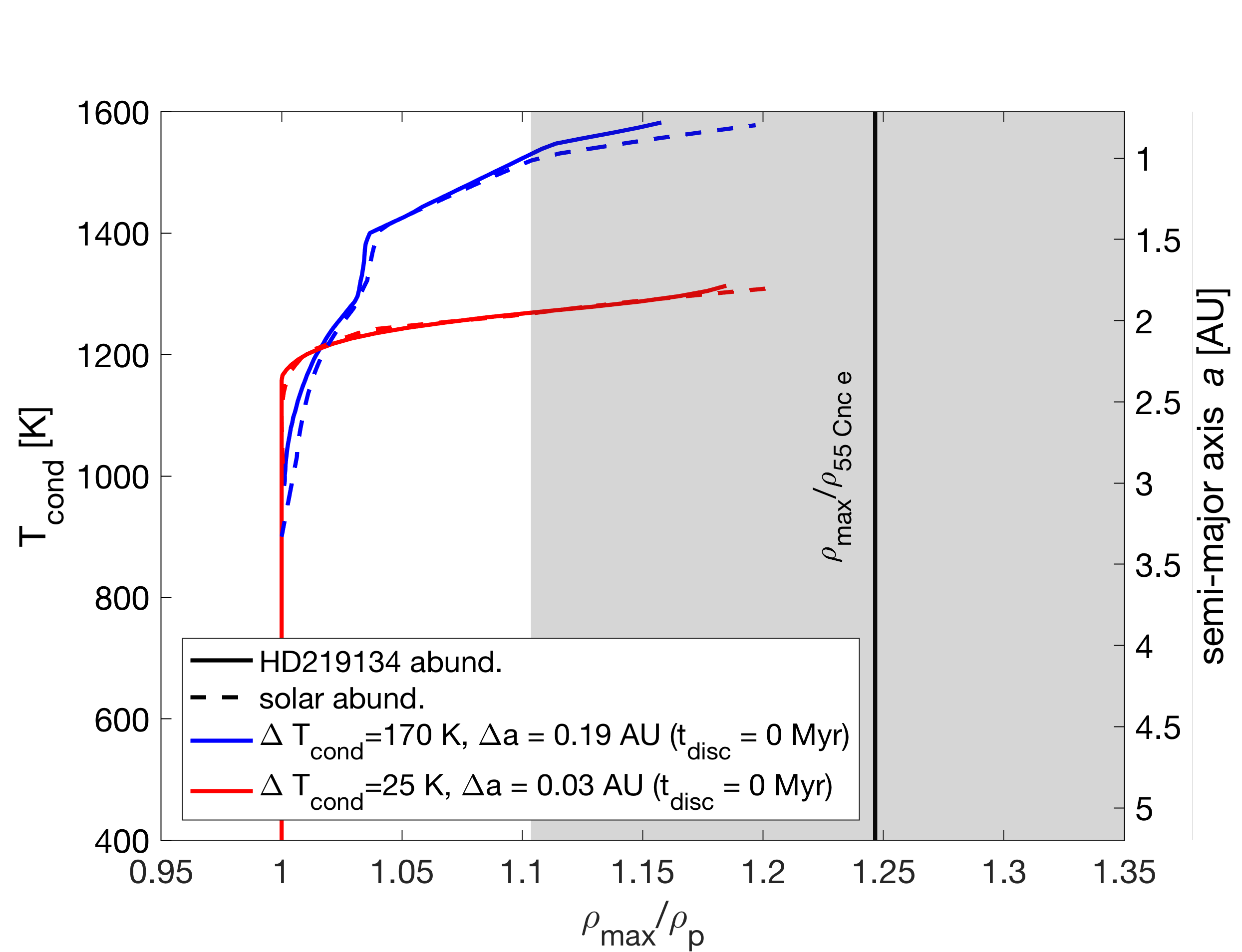}
    \caption{Possible variability in bulk density for a rocky planet plotted as the ratio $\rho_{\rm max}/\rho_{\rm p}$ as a function of temperature T$_{\rm cond}$ \rev{and semi-major axis $a$ (at $t_{\rm disc} = 0$\,Myr)}. Red and blue curves are adapted from cases of the Sun and HD219134, that show negligible differences due to their specific abundances. 55 Cnc e can be built from condensates that formed at temperatures T$_{\rm cond}$, because its density $\rho_{\rm 55 Cnc e}$ deviates from $\rho_{\rm max}$. $\rho_{\rm max}$ is the density that is realized for the majority of purely rocky planets which form at temperate conditions (<1200 K) around the star 55 Cnc. The specific stellar abundance of 55 Cnc is taken into account when calculating $\rho_{\rm max}$ and $\rho_{\rm 55 Cnc e}$ (see text). Grey bar denotes uncertainty of $\rho_{\rm max}/\rho_{\rm 55 Cnc e}$. }
    \label{fig:55cnc}
\end{figure}

Can we explain the low density of 55 Cnc e with a formation from high temperature condensates? In order to investigate this, we use the above discussed interior model (Section \ref{Sec:refr}). First, we compute the range of purely rocky interiors that fit the stellar abundance proxy while neglecting compositional variability that can occur in the PPD (Figure \ref{fig:MR_55cnce}). We use the stellar proxy stated in \citep{dorn2017bayesian} and assume a solar C/O ratio. In this case, bulk densities of 1.75-1.3 \DE can be reached within the measured 1-$\sigma$ uncertainty of the host star abundance (red area in Figure \ref{fig:MR_55cnce}), which is significantly denser than 55 Cnc e (with $\rho_{\rm 55 Cnc e} = 1.164 \pm 0.062 $\DE ). The range 1.75-1.3 \DE constitutes our stated  $\rho_{\rm max}$ in Figure \ref{fig:55cnc}.  The ratio of $\rho_{\rm 55 Cnc e}/\rho_{\rm max}$ is then compared to the variability in bulk density as inherited from chemically different planetesimals within the PPD of Sun-like stars in Figure \ref{fig:55cnc}. The red and blue curves are identical to those shown in Figure \ref{fig:plato}. 
A good fit to $\rho_{\rm 55 Cnc e}/\rho_{\rm max}$  is achieved by interiors built from high temperature condensates (> 1200 K) that are rich in Ca and Al. 

For 55 Cnc e's mass of 8.59 \ME, possible radii range from 1.7-1.88 \RE according to the measured uncertainty in the host star abundance. By allowing variation in rock composition as inherited from chemically different planetesimals, radii can reach 1.92 \RE for Ca and Al-rich interiors that are depleted in Fe. The planet radii can further increase to 1.94 - 1.95 \RE by the addition of a mineral 0.1 bar atmospheres (MgO, CaO, Na$_2$O). Such interior scenarios fit the observed radius of $R_{\rm 55 Cnc e} = 1.947 \pm 0.038 $\RE (Fig \ref{fig:MR_55cnce}). In addition, the potential escape of ionic calcium \citep{ridden2016search} would be consistent with our proposed interiors. In a follow-up study, we will investigate the case of 55 Cnc e in more detail to understand the possible exosphere species in case of a Ca and Al-rich interior.
 
We assumed solar C/O for 55 Cnc. Abundance estimates for the star vary considerably among different studies. A high C/O of 1.12 \citep{mena2010chemical} have been reported, which was subsequently followed by a lower estimate of 0.78 $\pm$ 0.08 \citep{teske2013carbon} allowing for carbon-rich interiors \citep{madhusudhan2012possible}. Recent analysis from \citet{brewer2016c} derive a C/O of 0.53 $\pm$ 0.054, almost identical to solar (0.54).

\section{WASP-47 e}
\label{Sec:wasp47}

\begin{figure}
	\includegraphics[width=1\columnwidth,trim = 1.5cm 0cm 2.5cm 0cm,clip]{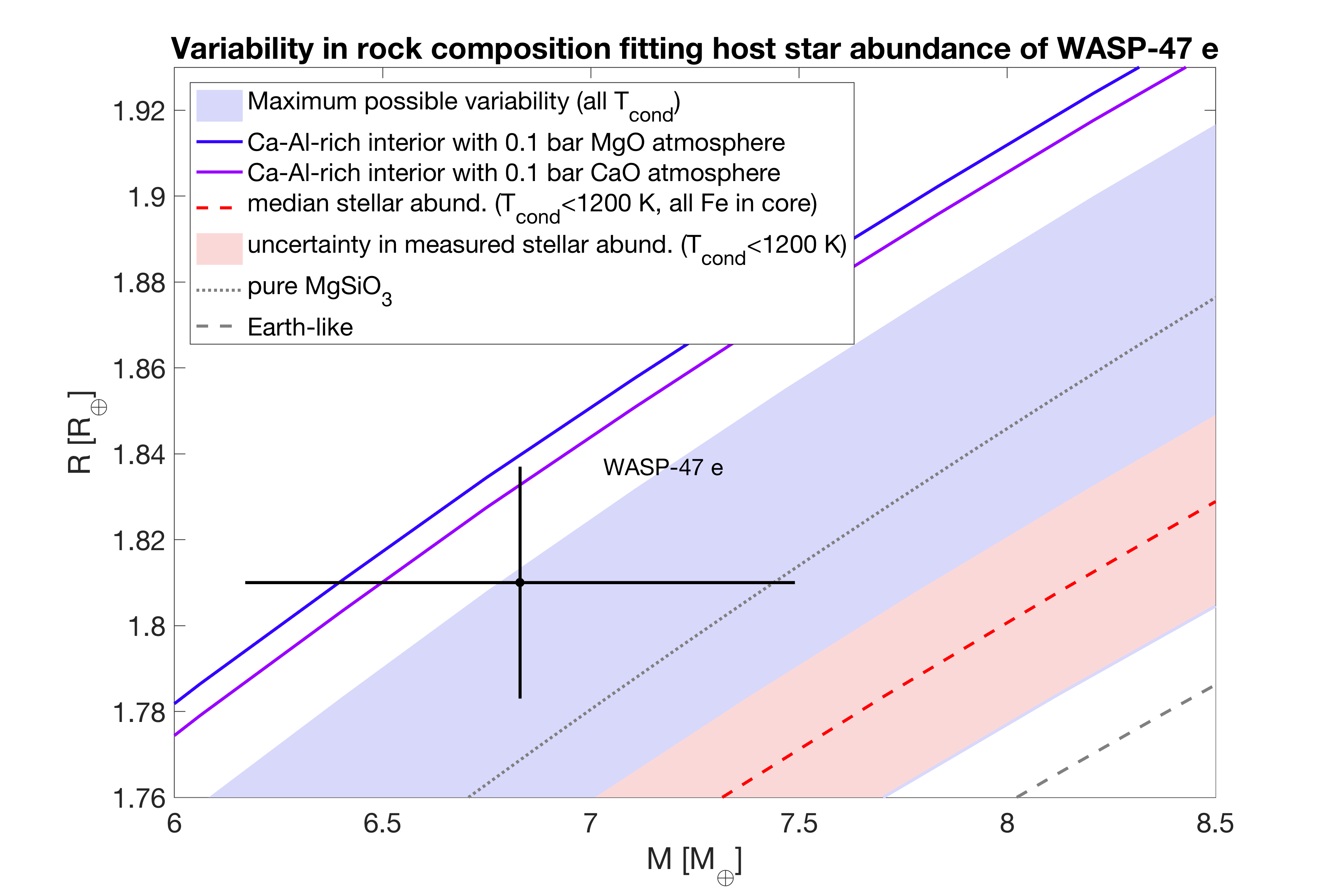}
    \caption{Mass-radius plot showing WASP-47 e compared to scaled idealized interiors. We assume a 1:1 compositional relationship (i.e., ratios of refractory elements) between the host star WASP-47 and the PPD in which WASP-47 e formed. The variability of purely rocky planets forming at different times and locations within the disc is highlighted by the blue area.
    Purely rocky interiors that are built from temperate condensates ($T_{\rm cond}<1200$K) follow the red area, which respects the uncertainty in measured stellar abundances. Interiors that fit the median stellar abundance follow the red curves. The solid colored curves assume the lowest density of a Ca and Al-rich composition plus a 0.1 bar mineral atmospheres.
    }
    \label{fig:wasp47}
\end{figure}

\begin{figure}
	\includegraphics[width=\columnwidth,trim = 0cm 0cm 0cm 1cm,clip]{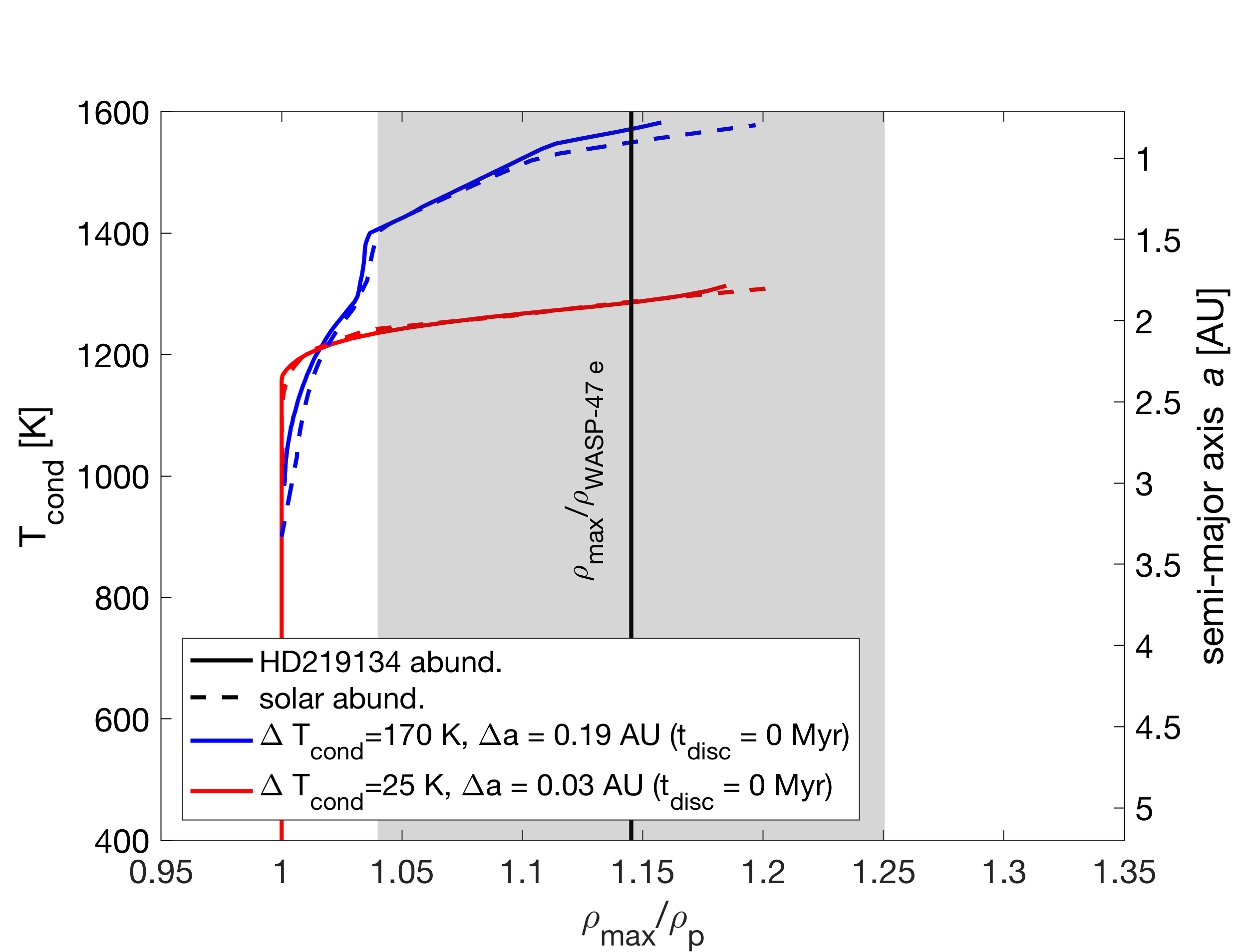}
    \caption{Possible variability in bulk density for a rocky planet plotted as the ratio $\rho_{\rm max}/\rho_{\rm p}$ as a function of temperature T$_{\rm cond}$ \rev{and semi-major axis $a$ (at $t_{\rm disc} = 0$\,Myr)}. Red and blue curves are adapted from cases of the Sun and HD219134, that show negligible differences due to their specific abundances.
    WASP-47 e can be built from condensates that formed at temperatures T$_{\rm cond}$, because its density $\rho_{\rm 55 Cnc e}$ deviates from $\rho_{\rm max}$. $\rho_{\rm max}$ is the density that is realized for the majority of purely rocky planets which form at temperate conditions (<1200 K) around the star WASP-47. The specific stellar abundance of WASP-47 is taken into account when calculating $\rho_{\rm max}$ and $\rho_{\rm WASP-47 e}$ (see text). Grey bar denotes uncertainty of $\rho_{\rm max}/\rho_{\rm WASP-47 e}$.}
    \label{fig:waspTrho}
\end{figure}

Similar to 55 Cnc e, the ultra-short period planet WASP-47 e has an unusually low density ($\rho_{\rm WASP-47 e} = 6.35 \pm 0.64$ g/cm$^3$ at $M_{\rm WASP-47 e} = 6.83 \pm 0.66$\ME) \citep{vanderburg2017precise}, that is inconsistent with Earth-like compositions (Figure \ref{fig:wasp47}). Given the stellar abundance estimates \citep{hellier2012seven} and their uncertainties, the possible range of bulk densities for rocky interiors  is 1.29-1.37 \DE, while neglecting compositional variability that can occur in high temperature regions of the PPD (Figure \ref{fig:wasp47}). The star has been estimated to be low in Fe and rich in Si (with mass ratios of Fe/Si$_{\rm WASP-47}$ = 1.12 and Mg/Si$_{\rm WASP-47}$ = 0.64), which is why the resulting density range of rocky interiors (1.29-1.37 \DE) is below an Earth-like composition. %\citet{hellier2012seven} do not provide the stellar Al abundance, which we assumed equal to those of  F-G-K stars of similar abundances in Fe, Mg, Si ([Al/H] $\sim$ 0.4).
The low bulk density can be explained by taking such chemical variability within the disc into account. In this case, the observed bulk density of $\rho_{\rm WASP-47 e} = 1.15 \pm 0.12$ \DE can be matched, suggesting the formation of WASP-47 e from high-temperature condensates.
In Figure \ref{fig:waspTrho}, the range 1.29-1.37 \DE constitute our stated  $\rho_{\rm max}$; $\rho_{\rm max}/\rho_{\rm WASP-47 e}$ is compared to the variability in bulk density as  inherited from chemically different planetesimals within the PPD of Sun-like stars. The red and blue curves (identical to those shown in Figure \ref{fig:plato}) can explain the planet properties only at high condensation temperatures.

Although the MgSiO$_3$ curve intersects the planetary data in Figure \ref{fig:wasp47}, it is an idealized composition that represents the lowest-density end-member of a purely rocky planet (dominated by Fe, Si, Mg) but unlikely exists in nature.

In summary, WASP-47 e can be explained by a Ca and Al-rich interior without thick gas layers. Given its high stellar irradiation, any gaseous envelope is subject to significant evaporative escape. So far, no characterization of the escaping atmosphere (exosphere) has been published. Given our Ca and Al-rich interior scenario, we predict the absence of escaping hydrogen originating from a H/He or water layer, but consider ionic calcium, silicon, magnesium, and maybe even aluminium to be possible in the exosphere. Which ions can be present in the exosphere is in part related to the vapour pressures, sputtering efficiencies, ionisation and escape rates of the different species. 
%Further investigations of WASP-47 e are needed to better understand its exotic interior.

\section{Further candidates}
\label{Sec:further}

\begin{figure}
	\includegraphics[width=0.95\columnwidth,trim = 0cm 0cm 1cm 0cm,clip]{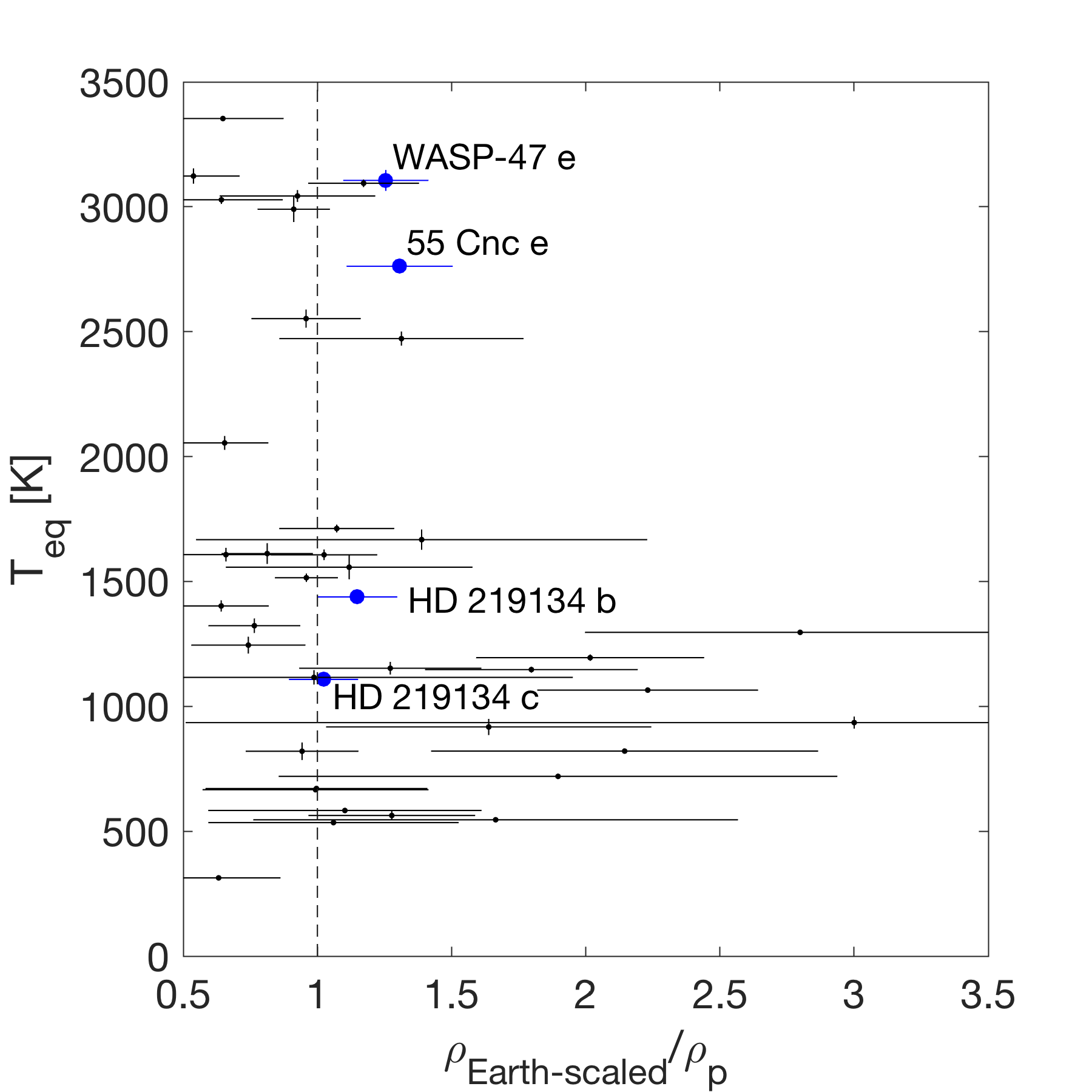}
    \caption{Variability in bulk density for confirmed planets plotted as the ratio $\rho_{\rm Earth-scaled}/\rho_{\rm p}$ as a function of equilibrium temperature T$_{\rm eq}$. $\rho_{\rm Earth-scaled}$ is the density that corresponds to an Earth-like composition for a given planet mass. Planets below 10 \ME and smaller than 2.2 \RE are shown for which mass and radius uncertainties are smaller than 30\% and 20\%, respectively. The highlighted planets are discussed in the main text, of which HD~219134~b, 55~Cnc~e, and WASP-47~e are candidates of a new class of Ca and Al-rich Super-Earths. }
    \label{fig:others}
\end{figure}

Future observations will show whether additional planetary systems can be found with Super-Earths that potentially formed from high-temperature condensates. If such Super-Earths exist, we expect to find them very close to their stars with densities 10-20 \% lower than Earth-like  compositions. Theoretically, Ca and Al-rich interiors are indeed  part of predicted planet populations from \citet{carter2012low,thiabaud2014stellar} who assume chemical equilibrium in PPDs. \citet{thiabaud2014stellar} find Ca and Al-rich planets at $< 0.3$ AU. They state bulk compositions of  40-50 wt\% Al, 7-9 wt\% Ca, which is similar to compositions investigated here.

\rev{An open question is how much mass is available in the innermost disc region to form massive Ca and Al-rich planets. Previous studies that employed N-body simulations find Ca and Al-rich planets of up to 1 \ME \citep{carter2012low} and 3 \ME \citep{thiabaud2014stellar}. Our proposed planet candidates are significantly more massive (up to 8.5 \ME). In order to form such massive planets close to the star, the disc surface density needs to be high enough in the innermost region. Disc properties are indeed not well constrained and adjustments in e.g., the total disc mass and the surface density gradient may allow to form massive Ca and Al-rich planets. However, there is an additional interesting aspect to the three candidates. All of them occur in systems with gas giants. This suggests that the gas giants may alter the disc structure in such a way that the formation of very close-in Super-Earths becomes possible. Although \citep{carter2012compositional} argue that Ca and Al-rich planets are extremely rare when migrating giant planets result in large-scale mixing of planetesimals, giant planets may also isolate planetesimal reservoirs by altering the disc structure as suggested for Jupiter \citep{alibert2018formation}.}

In Figure \ref{fig:others}, we plot confirmed Super-Earths and highlight the discussed planets.
%Plato science
PLATO \citep{rauer2014plato} will provide us with well-characterized planetary masses and radii that allows for 10\% precision on bulk density, which is \rev{necessary} to place strong constraints on the existence of other Ca and Al-rich interiors. \rev{However, mass and radius alone are insufficient to distinguish between interiors that are rich in either volatiles or Ca and Al. As demonstrated here, additional considerations of atmospheric escape (55 Cnc e, WASP-47 e) or constraints that stem from neighboring planets (HD219134 b) are required to conclude for the presence of Ca and Al-rich interiors. }

\section{Conclusions}
\label{conclusions}

We assumed that building blocks of rocky planets form from condensates of cooling PPDs. Very close to the star, temperatures in the gas disc are initially sufficiently high that most traditionally rocky species are vaporised. Theoretically, the building blocks that form at high temperatures
 ($>1200$ K) and in chemical equilibrium, can vary drastically in refractory element composition. A planet formed from these planetesimals can be rich in Ca and Al while being depleted in Fe. Here, we demonstrated that such compositional differences would be reflected in a lower bulk density of 10-20 \% compared to Earth-like compositions, even less than pure MgSiO$_3$. We have quantified the density variability of rocky planets as inherited from the chemical variability of planetesimals that formed in different temperature environments in the PPD. We demonstrated that there are at least three Super-Earths, HD219134 b, 55 Cnc e, and WASP-47 e for which Ca and Al-rich and core-free interiors can explain their observed properties.

\rev{Identifying Ca and Al-rich planets is impossible from bulk density alone. Additional constraints are necessary to rule out otherwise possible volatile-rich interiors that can have identical bulk densities. For our three studied candidates, these employed additional constraints differ. For the highly irradiated planets WASP-47 e and 55 Cnc e,} the thickness of volatile layers is limited given atmospheric escape and restricted outgassing. For HD219134 b, the existence of the Super-Earth HD219134 c of similar mass but higher bulk density imposes additional constraints on their interiors. For 55 Cnc e, the observed escape of ionic calcium additionally supports an interior rich in Ca, however, further investigations are required to understand the possible detection of sodium \citep{ridden2016search}.

In summary, HD219134 b, 55 Cnc e, and WASP-47 e are candidates of a new class of Super-Earths with interiors very different compared to the majority of Super-Earths or Earth-like compositions. Their interior dynamics, outgassing histories, atmosphere evolution, and magnetic fields may be fundamentally different than what we know from terrestrial Solar System planets.
We demonstrated that expected uncertainties provided by PLATO will allow us to study whether other planetary systems harbor equally exotic worlds that formed from high temperature condensates. 
\rev{On a popular science note, these worlds are rich in sapphires and rubies.}
%Their compositions allow for surfaces rich in sapphires!

\section*{Acknowledgements}

This work was supported by the Swiss National Foundation under grant PZ00P2\_174028. It was in part carried out within the frame of the National Center for Competence in Research Planets. The authors are also grateful to the Royal Society for funding this research through a Dorothy Hodgkin Fellowship and to the Science and Technology Facilities Council. We thank \rev{the reviewer Alexander Cridland for his valuable comments and also} thank James Connolly and Jonathan Fortney for insightful discussions.

%%%%%%%%%%%%%%%%%%%%%%%%%%%%%%%%%%%%%%%%%%%%%%%%%%



\bibliographystyle{mnras}
\bibliography{libary} % if your bibtex file is called example.bib

% \begin{thebibliography}{99}
% \bibitem[\protect\citeauthoryear{Author}{2012}]{Author2012}
% Author A.~N., 2013, Journal of Improbable Astronomy, 1, 1
% \bibitem[\protect\citeauthoryear{Others}{2013}]{Others2013}
% Others S., 2012, Journal of Interesting Stuff, 17, 198
% \end{thebibliography}

%%%%%%%%%%%%%%%%%%%%%%%%%%%%%%%%%%%%%%%%%%%%%%%%%%

%%%%%%%%%%%%%%%%% APPENDICES %%%%%%%%%%%%%%%%%%%%%

\appendix
\section{Mineral phases in Ca and Al-rich planets}
\label{phases}
\begin{figure}
	\includegraphics[width=\columnwidth,trim = 0.3cm 0cm 0.7cm 0cm,clip]{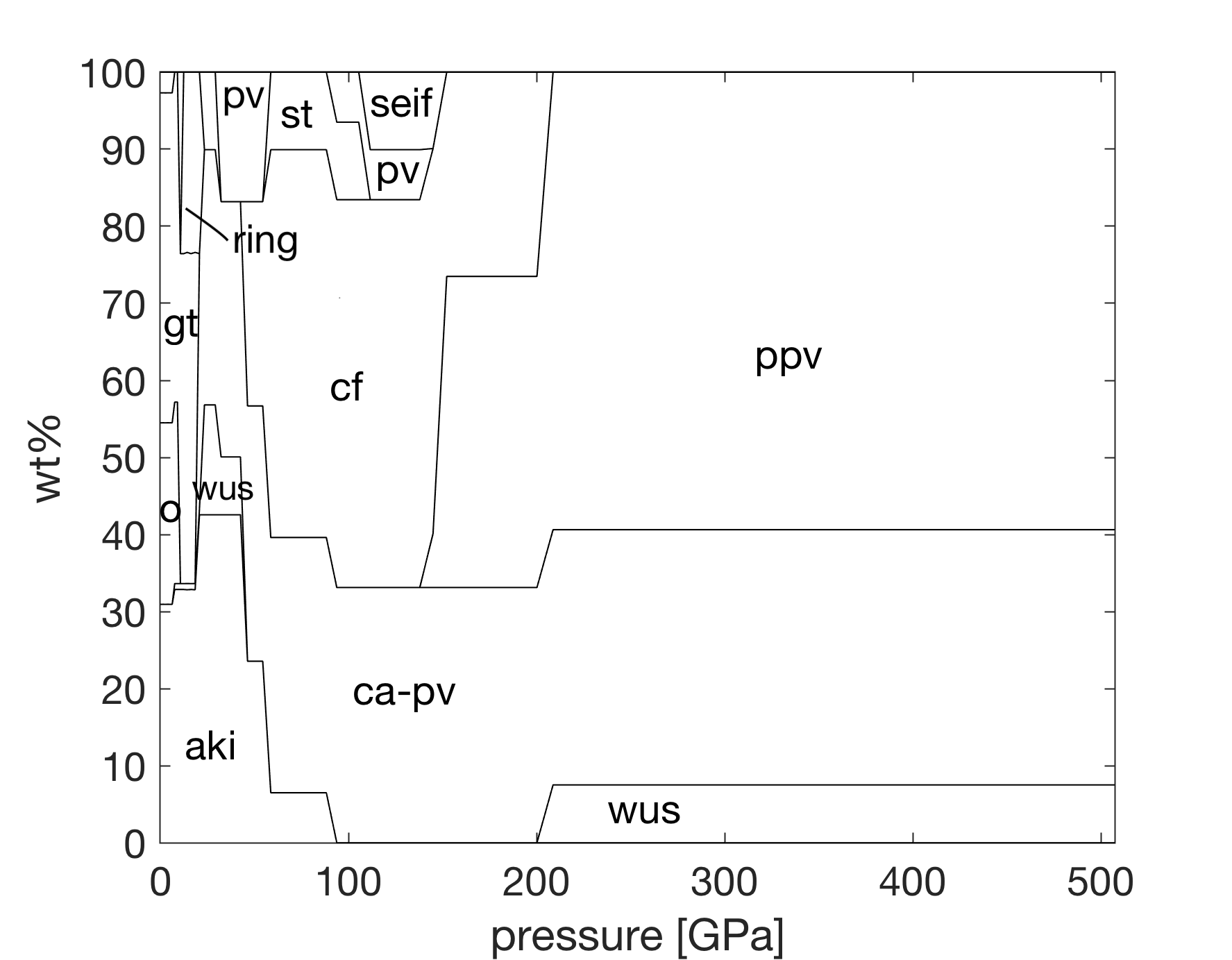}
    \caption{Variations in mineral phase proportions computed for a planet of 5 \ME with a bulk composition of CaO (16\%), FeO ($\ll0.1$\%), MgO (14\%), Al$_2$O$_3$ (43\%), SiO$_2$ (27\%), Na$_2$O ($\ll0.1$\%). This is one realization of a Ca and Al-rich interior. Indicated mineral phases are akimotoite (aki), calcium-perovskite (ca-pv), calcium-ferrite (cf), wustite (wus), perovskite (pv), post-perovskite (ppv), olivine (o), garnet(gt), ringwoodite (ring), stishovite (st), seifertite (seif).  }
    \label{fig:phases}
\end{figure}

For one realization of a rocky Ca and Al-rich interior, we provide the computed mineral phase proportions as a function of pressure Figure \ref{fig:phases}. These calculations are done with perple\_X \citep{connolly2009geodynamic} using the thermodynamic data of \citet{stixrude2011thermodynamics}. The thermodynamic data covers pressure conditions as we find them in Earth, here we rely on the extrapolation of these data to significantly higher pressures. Largest uncertainties in the thermodynamic data are expected for the aluminium end-members and especially for the calcium-ferrite phase.

\section{Evaporative loss}
\label{App:massloss}
As introduced by \citet{dorn2018secondary}, there is a minimum threshold thickness for a primordial atmosphere $\Delta R_{\rm th}$ below which the life-time due to evaporative loss is too short to be observable. The minimum thickness corresponds to a minimum mass of gas. While \citep{dorn2018secondary} estimate this minimum mass by the amount of gas that is lost during the age of the star, we use the amount of gas that is lost over 100 Myr, which is less than one percent of the stellar age of HD219134 of 12.9 Gyr \citep{takeda2007structure}.

Here, the minimum threshold-thickness $\Delta R_{\rm th}$ is defined as
\begin{equation}
\label{Eq:threshold}
\Delta R_{\rm th}  =H \ln{\left(\frac{P_b}{P_{\rm photo}}\right)} \approx H \ln{\left(\frac{g_{\rm surf} \dot{m}_{\rm gas} 100 {\rm Myr}/ 4 \pi R_{\rm p}^2 }{P_{\rm photo}}\right)} \,,
\end{equation}
where the pressure scale height $H$ equals the universal gas constant $R^*$ times equilibrium temperature $T_{\rm eq}$ divided by mean molecular weight $\mu$ and surface gravity $g_{\rm surf}$. $P_b$ corresponds to the pressure that the total amount of lost gas (over 100 Myr) would impose.
 The energy-limited gas loss rate $\dot{m}_{\rm gas}$ is calculated following \citet{lopez2017born}:
\begin{equation}
\label{eq:masslossrate}
\dot{m}_{\rm gas} = - \frac{\eta \pi F_{\rm XUV} R_{\rm base}^3}{G M_{\rm p} K_{\rm tide}}\,,
\end{equation}
where $\eta$ is the photoevaporation efficiency or the fraction of incident XUV energy that removes gas. $F_{\rm XUV}$ is the XUV flux of the star ($4 \times 10^{26}$ erg/s as estimated by \citet{porto2006astrobiologically}), $G$ is the gravitational constant, $M_{\rm p}$ is planet mass, $K_{\rm tide}$ accounts for contributions of tidal forces to the potential energy and is set to unity for simplicity. The dependency of $\dot{m}_{\rm gas}$ on the mean molecular weight $\mu$ is introduced through $H$ by $R_{\rm base}$  which is the planetary radii at the base of the XUV photosphere and is approximated by 
\begin{equation}
R_{\rm base} = R_{\rm p} + H \ln{\frac{P_{\rm photo}}{P_{\rm base}}}\,,
\end{equation}
where $R_{\rm p}$ is planet radii at $P_{\rm photo}$ that is the pressure  at the optical photosphere in a transit geometry. $P_{\rm photo}$ is approximated by $\nicefrac{g_{\rm surf}}{\kappa} \sqrt{H / (2\pi R_{\rm p})}$, with $\kappa = 0.1$ cm$^2$ g$^{-1}$  \citep{freedman2014gaseous} and is about 1 mbar for both planets.

The pressure at the XUV photosphere $P_{\rm base}$ is approximated by $m_{\rm u} \mu g_{\rm surf} / \sigma_{\nu 0} $, where $m_{\rm u}$ is the atomic mass constant and $\sigma_{\nu 0} = 6 \times 10^{-18} (h\nu_0/E_i)^{-3}$ cm$^{2}$ with a photon energy of $h\nu_0 = 20$ eV \citep{murray2009atmospheric} and the ionization energy $E_i$ for hydrogen of 13.6 eV. $P_{\rm base}$ is about 1 nbar.

\section{Further details on the PPD model}
\label{discmodel}
In the following section $t_{\rm disc}$ is as defined in the main body of the paper as the time in the PPD phase. $a$ is the radial distance from the host star as defined in the main body of the paper. $M_{0}$ is the mass of the PPD which is related to the mass of the host star via $M_{0}= 0.1M_{*}$. $s_{0}$ defined as the initial radial extent of the PPD and is kept at 33\,AU in accordance with \cite{Chambers2009}. All other constants are represented by their standard symbols.

The inner viscous evaporating region has a surface density given by 
\begin{equation}
\Sigma(a,t_{\rm disc}) = \Sigma_{evap} \left(\frac{a}{s_{0}}\right)^{-\frac{24}{19}}  \left( 1 + \frac{t_{\rm disc}}{\tau_{vis}}\right)^{-\frac{17}{16}}
\end{equation}
where 
\begin{equation}
\Sigma_{evap} = \Sigma_{vis} \left(\frac{T_{vis}}{T_{e}}\right)^{\frac{14}{19}}
\end{equation}
$ T_{e} = 1380\,K$
The opacity in the inner viscous evaporating region follows the power law described in \cite{Stepinski1998}.
\par
\noindent The temperature in the viscous evaporating inner region is given by
\begin{equation}
T(a,t_{\rm disc}) = T_{vis}^{\frac{5}{19}} T_{e}^{\frac{14}{19}}  \left(\frac{a}{s_{0}}\right)^{-\frac{9}{38}}  \left( 1 + \frac{t_{\rm disc}}{\tau_{vis}}\right)^{-\frac{1}{8}}
\end{equation}
and the transition radius to the intermediate viscous region is
\begin{equation}
r_{e} (t_{\rm disc}) = s_{0} \left(\frac{\Sigma_{evap}}{\Sigma_{vis}}\right)^{\frac{95}{63}} \left( 1 + \frac{t_{\rm disc}}{\tau_{vis}}\right)^{-\frac{19}{36}}
\end{equation}
\noindent The surface density in the intermediate viscous region is
\begin{equation}
\Sigma(a,t_{\rm disc}) = \Sigma_{vis} \left(\frac{a}{s_{0}}\right)^{-\frac{3}{5}}  \left( 1 + \frac{t_{\rm disc}}{\tau_{vis}}\right)^{-\frac{57}{80}}
\end{equation}
where
\begin{equation}
\Sigma_{vis} = \frac{7M_{0}}{10 \pi s_{0}^{2}}
\end{equation}
The temperature in the intermediate viscous region is
\begin{equation}
T(a,t_{\rm disc}) = T_{vis} \left(\frac{a}{s_{0}}\right)^{-\frac{9}{10}}  \left( 1 + \frac{t_{\rm disc}}{\tau_{vis}}\right)^{-\frac{19}{40}}
\end{equation}
where 
\begin{equation}
T_{vis} = \left(\frac{27 \kappa_{0}}{64 \sigma}\right)^{\frac{1}{3}}  \left(\frac{ \alpha \gamma k }{\mu m_{H}}\right)^{\frac{1}{3}}   \left(\frac{7 M_{0}}{10 \pi  s_{0}^{2}}\right)^{\frac{2}{3}}  \left(\frac{G M_{*} }{s_{0}^{3}}\right)^{\frac{1}{6}} 
\end{equation}
and
\begin{equation}
\tau_{vis} = \frac{1}{16 \pi} \frac{\mu m_H \Omega_{0} M_{0}}{\alpha \gamma k \Sigma_{vis} T_{vis}}
\end{equation}
and the transition radius between the intermediate viscous region and the outer irradiated region is
\begin{equation}
r_{t} (t_{\rm disc}) = s_{0} \left(\frac{\Sigma_{rad}}{\Sigma_{vis}}\right)^{\frac{70}{33}} \left( 1 + \frac{t_{\rm disc}}{\tau_{vis}}\right)^{-\frac{133}{132}}
\end{equation}
\noindent The surface density in the outer irradiated region is
\begin{equation}
\Sigma(a,t_{\rm disc}) = \Sigma_{rad} \left(\frac{a}{s_{0}}\right)^{-\frac{15}{14}}  \left( 1 + \frac{t_{\rm disc}}{\tau_{vis}}\right)^{-\frac{19}{16}}
\end{equation}
where 
\begin{equation}
\Sigma_{rad} = \Sigma_{vis} \frac{T_{vis}}{T_{rad}}
\end{equation}
and
\begin{equation}
T_{rad} = \left(\frac{4}{7}\right)^{\frac{1}{4}}  \left(\frac{ T_{*} R_{*} k }{G M_{*} \mu m_{H}}\right)^{\frac{1}{7}}  \left(\frac{R_{*}}{s_{0}}\right)^{\frac{3}{7}}  T_{*}
\end{equation}
and the temperature in the outer irradiated region is
\begin{equation}
T(a) = T_{rad} \left(\frac{a}{s_{0}}\right)^{-\frac{3}{7}}  
\end{equation}
To convert the surface density profile into a pressure profile we have assumed the disc is an ideal gas with a Gaussian density profile.  The surface density is converted into a pressure as follows:
as
\begin{equation}
P = \frac{k \rho T}{\mu m_{H}}
\end{equation}
and
\begin{equation}
\int \rho \,dz = \Sigma 
\end{equation}
and we assume that
\begin{equation}
\rho = \rho_{0} e^{-\frac{z^{2}}{2H^{2}}}
\end{equation}
we therefore find that
\begin{equation}
\Sigma = \rho_{0} \sqrt{2H^{2} \pi }
\end{equation}
hence the pressure at the midplane is
\begin{equation}
P = \frac{k \Sigma T}{\mu m_{H} H \sqrt{2 \pi} }
\end{equation}
Using the standard formulae
\begin{equation}
c^{2}_{s} = \frac{kT}{\mu m_{H}}
\end{equation}
 \begin{equation}
H = \frac{c_{s}}{\Omega}
\end{equation}
\begin{equation}
\Omega = \sqrt{\frac{G M_{*}}{a^{3}}}
\end{equation}
we find that the relationship between the pressure profile and the surface density profile is
\begin{equation}
P = \sqrt{\frac{G M_{*}  k\Sigma^{2}T}{2 \pi \mu m_{H} a^{3}}}
\end{equation}

\begin{figure}
	\includegraphics[width=\columnwidth]{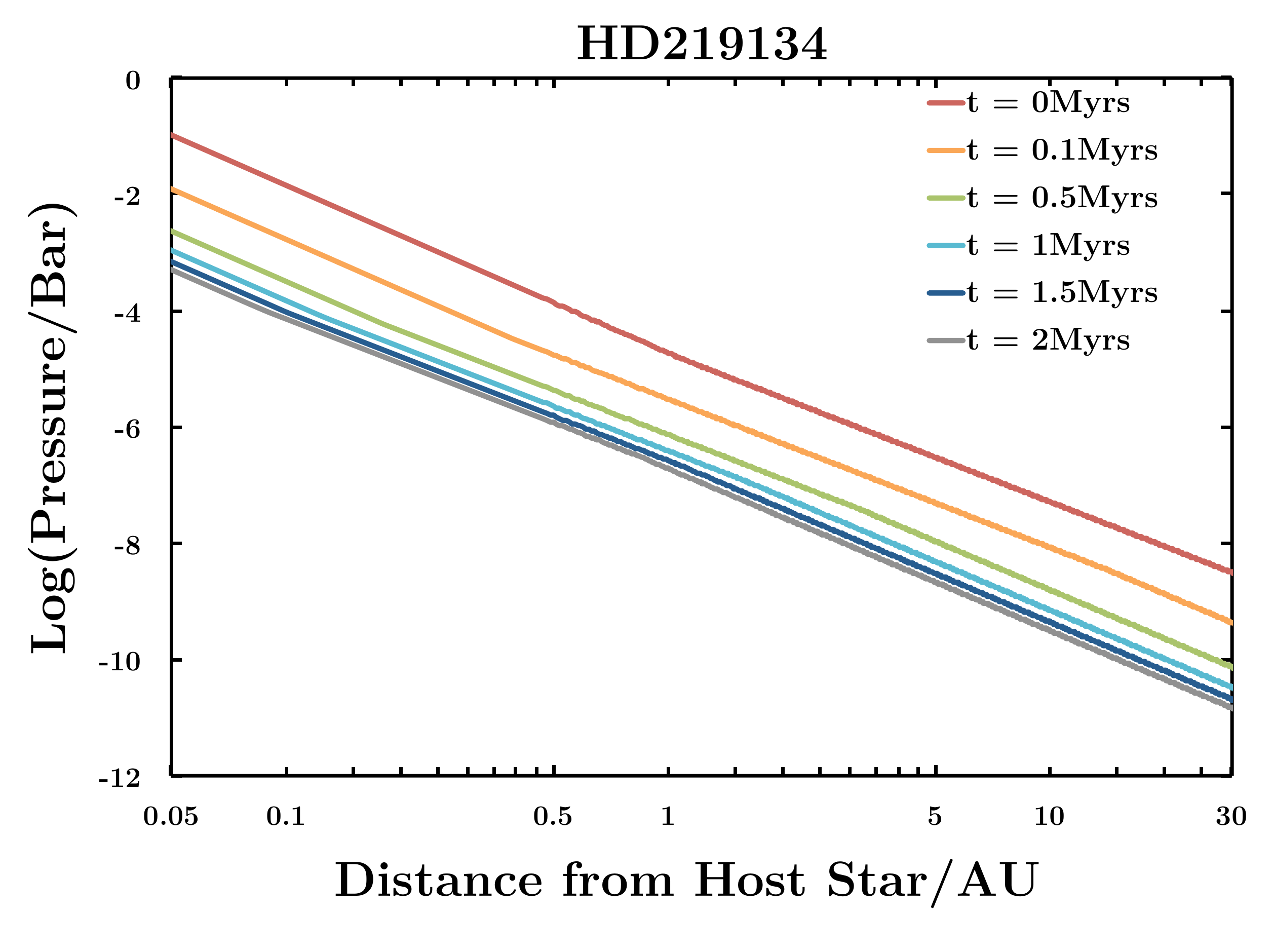}\\
    \includegraphics[width=\columnwidth]{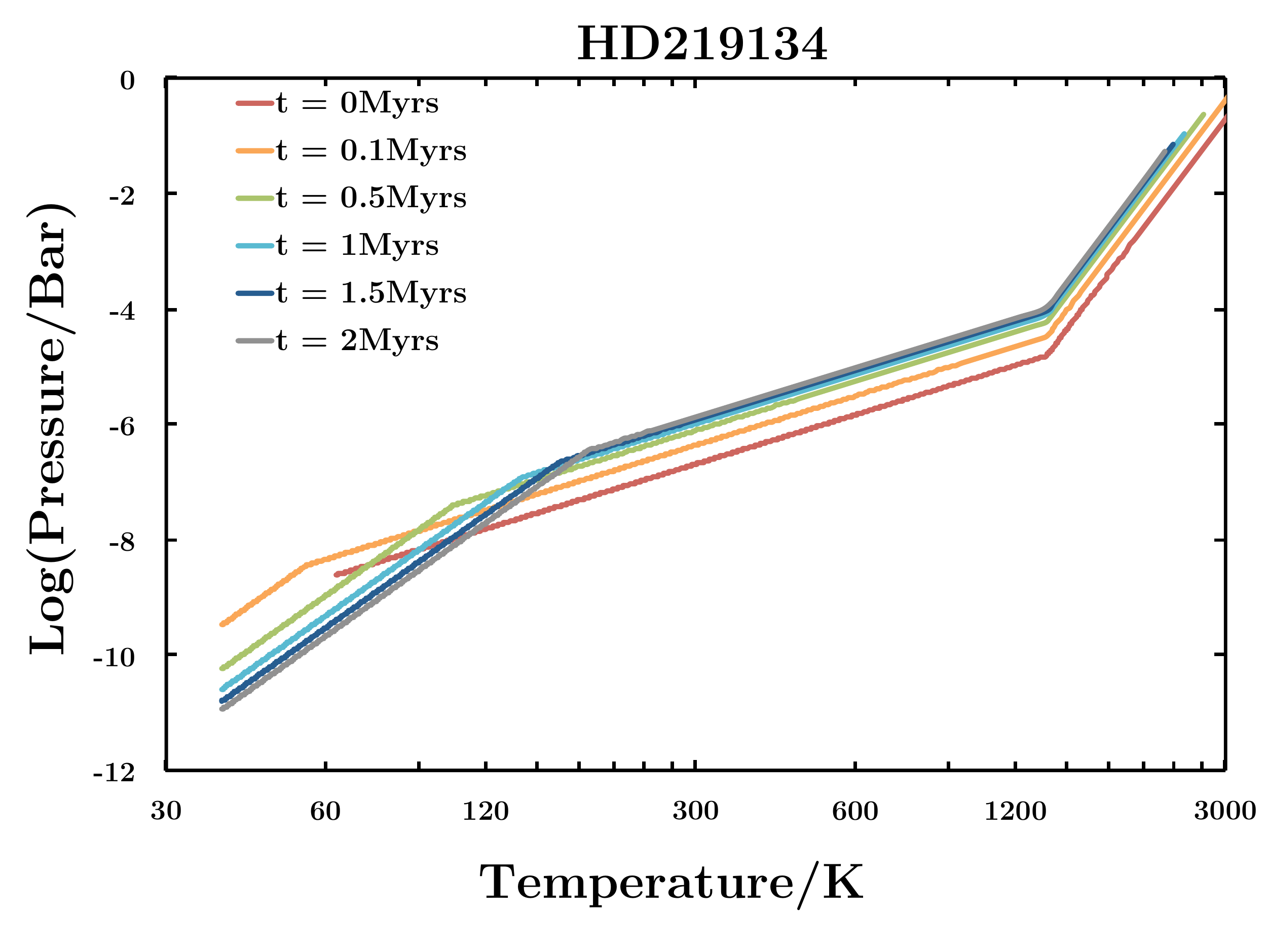}
    \caption{(Upper panel) pressure-temperature and (lower panel) pressure-radial distance space mapped out by the HD219134-like PPD as a function of time.}
    \label{fig:HDPT}
\end{figure}

%%%%%%%%%%%%%%%%%%%%%%%%%%%%%%%%%%%%%%%%%%%%%%%%%%

% Don't change these lines
\bsp	% typesetting comment
\label{lastpage}
\end{document}